\documentclass[journal]{IEEEtran}

\usepackage{cite}

\usepackage[cmex10]{amsmath}
\usepackage{nccmath}

\usepackage{array}
\usepackage{graphicx}
\usepackage{epstopdf}

\usepackage{gensymb}

\usepackage{url}

\usepackage{comment}

\begin{document}
\title{Characterization of Silicon Photomultipliers for nEXO}
%
%
%
\author{
I.~Ostrovskiy,
F.~Retiere,
D.~Auty,
J.~Dalmasson,
T.~Didberidze,
R.~DeVoe,
G.~Gratta,
L.~Huth,
L.~James,
L.~Lupin-Jimenez,
N.~Ohmart,
A.~Piepke%
\thanks{I. Ostrovskiy, R. DeVoe, G Gratta, and L. Lupin-Jimenez are with Physics Department, Stanford University, Stanford, California 94305 USA. (e-mail: ostrov@stanford.edu)}
\thanks{F. Retiere, L. Huth, L. James, and N. Ohmert are with TRIUMF, Vancouver, British Columbia V6T 2A3, Canada}
\thanks{D. Auty, T. Didberidze, and A. Piepke, are with Department of Physics and Astronomy, University of Alabama, Tuscaloosa, Alabama 35487, USA}%
\thanks{D. Auty is now at University of Alberta, Edmonton, Alberta, Canada}%
\thanks{J. Dalmasson is with Physics Department, University of Massachusetts, Amherst, Massachusetts 01003, USA}%
\thanks{
This work was supported by the U.S. Department of Energy Offices of High Energy and Nuclear Physics, 
Natural Sciences and Engineering Research Council of Canada, and TRIUMF Science Technology Department.}%
}

\maketitle

\begin{abstract}
Silicon Photomultipliers (SiPMs) are attractive candidates for light detectors for next generation liquid
xenon double-beta decay experiments, like nEXO (next Enriched Xenon Observatory). In this paper we discuss the requirements that the SiPMs must
 satisfy in order to be suitable for nEXO and similar experiments, describe the two test setups operated by the nEXO
collaboration, and present the results of characterization of SiPMs from several vendors.
In particular, we find that the photon detection efficiency at the peak of xenon scintillation light emission (175-178 nm) approaches  the nEXO requirements for tested FBK and Hamamatsu devices. Additionally, the nEXO collaboration performed radio-assay of several grams of bare FBK devices using neutron activation analysis, indicating levels of $^{40}$K, $^{232}$Th, and $^{238}$U of the order of $<$0.15, (6.9$\cdot$10$^{-4}$ - 1.3$\cdot$10$^{-2}$), and $<$0.11 mBq/kg, respectively.
\end{abstract}


%
\IEEEpeerreviewmaketitle

\section{Introduction}

\IEEEPARstart{S}{earch} for neutrinoless double-beta decay (DBD) is an
active field of research with important implications for nuclear and particle physics. nEXO is a planned 5-ton liquid xenon detector aimed to probe the effective Majorana neutrino mass in the inverted mass hierarchy region.
nEXO benefits from experience gained with the successful EXO-200 detector~\cite{Albert:2014b}, which utilizes $\sim$200 kg of liquid xenon kept at a nominal pressure of $\simeq$147 kPa and a temperature of $\simeq$167 K.
To achieve maximum sensitivity, the nEXO detector is required to have an energy resolution of $\frac{\sigma_E}{E}\sim$1\% at 2.458 MeV~\cite{QVal:1998} (Q value of DBD in $^{136}$Xe). 
The experiment further needs to achieve unprecedented levels of background in order to be able to detect a weak effect. A detector containing the largest physically or fiscally feasible amount of the decaying substance maximizes the decay rate.

In the EXO-200 experiment, the resolution is limited by the noise in the photodetector
channels. EXO-200 uses Large Area Avalanche Photo-diodes (LAAPDs) operated at a gain of
$\sim$200 and an equivalent noise of $\sim$10 photoelectrons per readout channel. Alternative light sensors that are gaining popularity
in recent years are Silicon Photomultipliers (SiPMs). SiPMs are photodetectors under investigation for use in nEXO because of
their high gain and expected low radioactive content. Due to large gains, on the order of 10$^6$, SiPMs have the potential to yield
much higher signal-to-noise ratios, to the point that it is not unreasonable to expect noise below 1 photoelectron equivalent. SiPM devices that are currently available need to exhibit the following characteristics:
\subsubsection{High photon detection efficiency (PDE) at 175-178 nm (peak of Xe scintillation light emission)}
The aim for overall efficiency for detecting the scintillation photons produced in liquid Xenon is 10\%. 
Simulations show that this efficiency is achievable if the sensors' PDE is at least 15\%.
Note that this includes loss of light due to reflection off the SiPM's surface, which is large ($\sim$50\%) when SiPMs are submerged in liquid xenon due to the large mismatch in indices of refraction at 175 nm between silicon ($n\sim0.8+1.9i$~\cite{n_silicon}) and xenon ($n=1.7$~\cite{n_xenon}). As a first step of the photodetector R\&D we perform measurements in vacuum in order to more easily test different  
photodetectors. Once devices passing our first level tests (described in this paper) have been identified, their functionality in liquid xenon will be verified and reflectivity of the photodetectors will be measured as a function of angle.

\subsubsection{High Radiopurity}
We demand that the content of the photodetectors has to be a sub-dominant contributor to the background budget, compared to other sources (e.g., the detector vessel). As a guide, we currently require that the photodetectors contain no more than $\sim$ppt levels of $^{238}$U and $^{232}$Th, which corresponds to specific activities on the order of $\sim$10 $\mu$Bq/kg. In practice, any solution that involves packaging material added to the bare device, especially ceramic, is disadvantageous, compared to using bare detectors.

\subsubsection{Small values of photodetector nuisance parameters}
Dark noise pulses start contributing to the dead time if their rate exceeds $\sim$200 MHz when summing over all SiPMs. Assuming 4 m$^2$ photo-coverage, the dark noise rate limit is 50 MHz/m$^2$ or 50 Hz/mm$^2$. This dark noise rate is to be achieved at liquid xenon temperatures. Correlated avalanches in SiPMs are caused by cross-talk (CT) and after-pulses (AP). They start contributing to the energy resolution if the fraction of additional (correlated) avalanches created by one avalanche exceeds 0.2.

\subsubsection{Good photodetector electrical properties}
Scintillation light is emitted isotropically, hence the spatial distribution of the light does not need to be sampled finely. Decreasing the number of readout channels also allows us to minimize the number of required feedthroughs. The size of single photodetector elements should be at least 1$\times$1 cm$^2$ (the current baseline design for nEXO assumes a unit readout cell of 5 cm$^2$). Larger areas, up to 20$\times$20 cm$^2$, are preferable but are limited by the electronics noise that scales with the capacitance per channel. Therefore, low capacitance per unit area is desirable (e.g. $<$50 pF/mm$^2$)~\cite{Lorenzo:2014}.

\subsubsection{Several ns timing resolution}
Good timing resolution is not essential for nEXO. Several ns single photon timing resolution would be sufficient. Nevertheless, single photon timing resolution might prove useful for other applications, by allowing better pulse shape discrimination and position reconstruction with light only using time of flight.

\subsubsection{Compatibility with liquid xenon}
Very long electron life time ($\sim$10 ms) must be achieved in nEXO, due to the long drift distances in a large detector, which requires very low levels of electron absorbing contaminants. Therefore, the photodetectors and associated ancillary mechanical and electrical interfaces will have to be tested in liquid xenon together with the passive optical elements to ensure acceptable levels of out-gassing. Furthermore, operation of the complete light detection system must be demonstrated in liquid xenon. In practice, any solution that involves wavelength shifters, which are commonly used to enhance PDE in the VUV region, poses additional risks, compared to the operation of bare devices.

Table~\ref{tab:reqs} summarizes the required SiPM parameters.

\begin{table}[t]
\caption{Summary of SiPM parameters required by nEXO. Parameters in italic are preferable, but not mandatory.}
\centering
\begin{tabular}{l@{\hskip 0.2in} l }
\hline\hline
\textbf{Parameter} & \textbf{Value} \\
\hline
\parbox[t]{5.5cm}{Photo-detection efficiency at 175-178 nm \\(without anti-reflective coating in gas/vacuum)} &  \(\geq\)15\% 	\\
&\\
\hline
Radio-purity: & \\
\hspace{0.5cm}$^{232}$Th and $^{238}$U & $<$10 $\mu$Bq/kg \\   
&\\
\hline
Dark noise rate at -100\celsius & $\leq$50 Hz/mm$^2$\\
&\\
\hline
After-pulse and cross-talk probability & $\leq$20\%\\
&\\
\hline
Single photodetector active area & $\geq$1 cm$^2$\\
&\\
\hline
\textit{Gain fluctuations and electronic noise} & $\leq$0.1 p.e.\\
&\\
\hline
\textit{Single photon timing resolution} & $<$10 ns\\
\hline\hline
\end{tabular}
\label{tab:reqs}
\end{table}
 
\begin{figure}[htbp]
 \centering
 \includegraphics[width=0.23\textwidth]{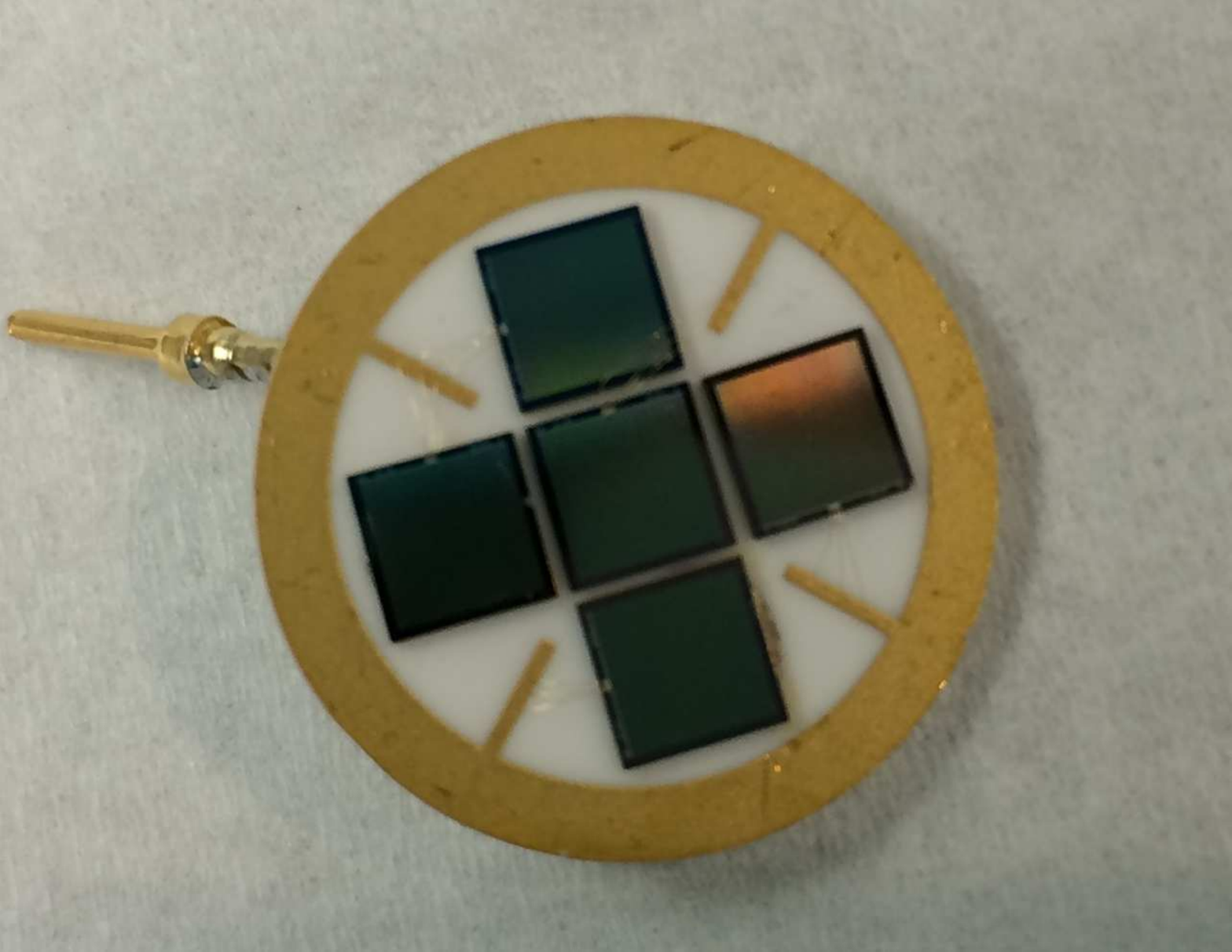}
 \includegraphics[width=0.23\textwidth]{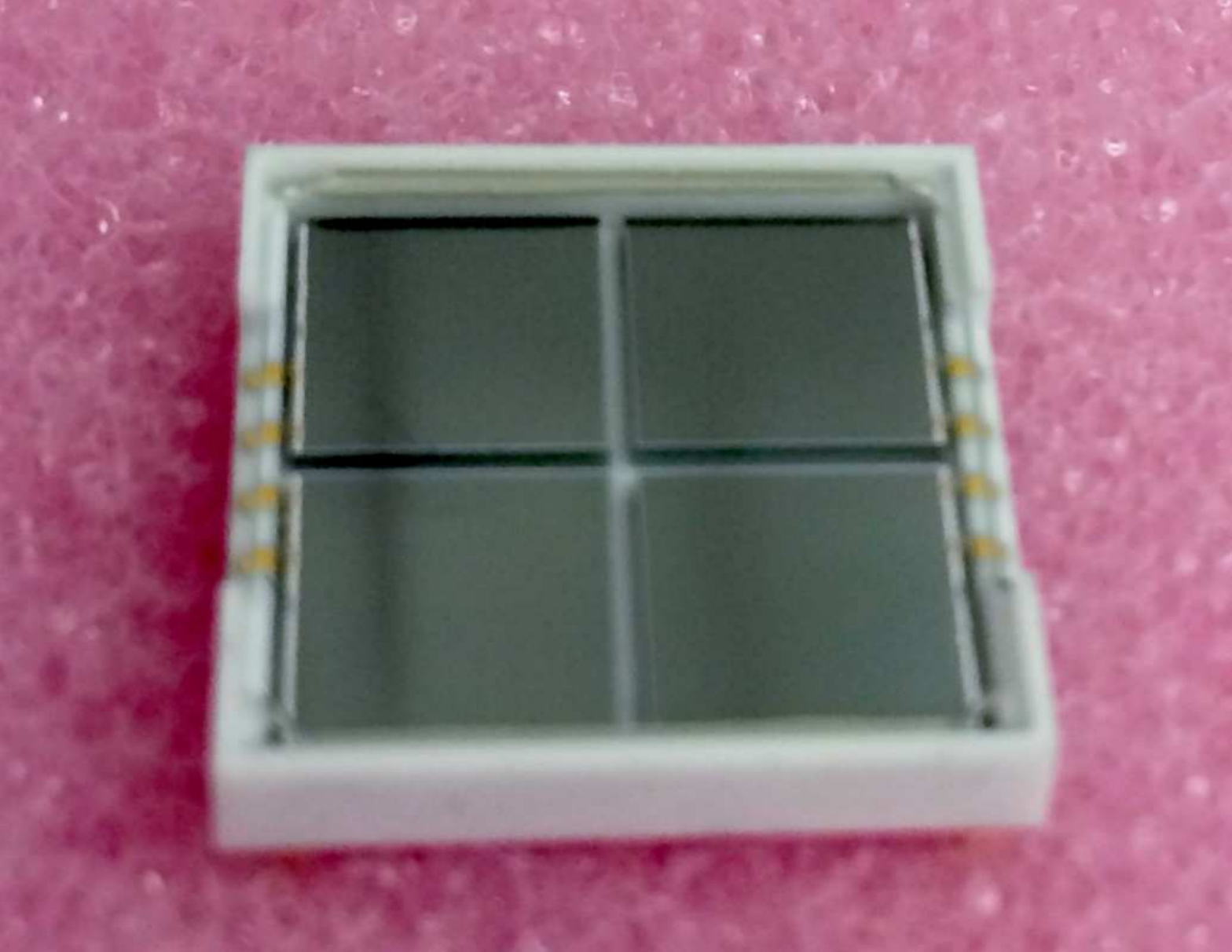}
  \caption{Examples of SiPMs used in this study. Five 4$\times$4 mm$^2$ FBK-2010 SiPM devices connected in parallel (left). The FBK-2010 devices were developed in 2010 based on their original n$^+$/p technology and feature thin silicon oxide window to facilitate UV light transmittance. Four 6$\times$6 mm$^2$ Hamamatsu-VUV devices connected in parallel (right) developed for the MEG experiment, optimized for UV detection.}
 \label{fig:sipms}
\end{figure}

The challenges listed above are addressed in close collaboration with SiPM manufacturer Fondazione Bruno Kessler (FBK). In addition,
SiPMs from other manufacturers, such as  Hamamatsu, KETEK, Excelitas, Zecotek, and RMD, are under investigation too.
We currently operate two setups for testing and characterization of SiPMs, allowing measurements of key
parameters at cryogenic temperatures. 
These two setups are described in the following sections along with current results obtained with the FBK, Hamamatsu, and KETEK
samples. Details on the devices used in this study are listed in Table~\ref{tab:devices}. The pictures of representative SiPM samples
are shown Fig~\ref{fig:sipms}.

\begin{table}[t]
\caption{SiPMs used in this study. Devices for which PDE measurements have been performed are given in Italic.}
\centering
\begin{tabular}{l c  c  c }
\hline\hline
\textbf{Producer} & \textbf{UV sensitive} & \textbf{Bare device} & \textbf{Comments} \\
\hline
\textit{FBK} & 	Yes & Yes & \parbox[t]{2.95cm}{Developed in 2010 based on the original n$^+$/p FBK SiPM technology~\cite{fbk:2013}. This type will be referred as FBK-2010 in the text}\\
\hline
\textit{Hamamatsu} & Yes & No &\parbox[t]{2.95cm}{Developed in 2013 for the MEG experiment~\cite{MEG:2014}.
This type will be referred as Hamamatsu-VUV in the text}\\
\hline
KETEK & No & No & \\
\hline
FBK & Yes & Yes & \parbox[t]{2.95cm}{Developed in 2013 based on the "RGB" FBK technology~\cite{fbk:2013}. 
This type will be referred as FBK-2013 in the text}\\
\hline
Hamamatsu & No & No &\parbox[t]{2.95cm}{Similar technology to MEG, no UV enhancements. This device
will be referred to as Hamamatsu-VIS in the text}\\
\hline\hline
\end{tabular}
\label{tab:devices}
\end{table}

\section{Test setup for PDE measurement}
\begin{figure}[htbp]
 \centering
 \includegraphics[width=0.45\textwidth]{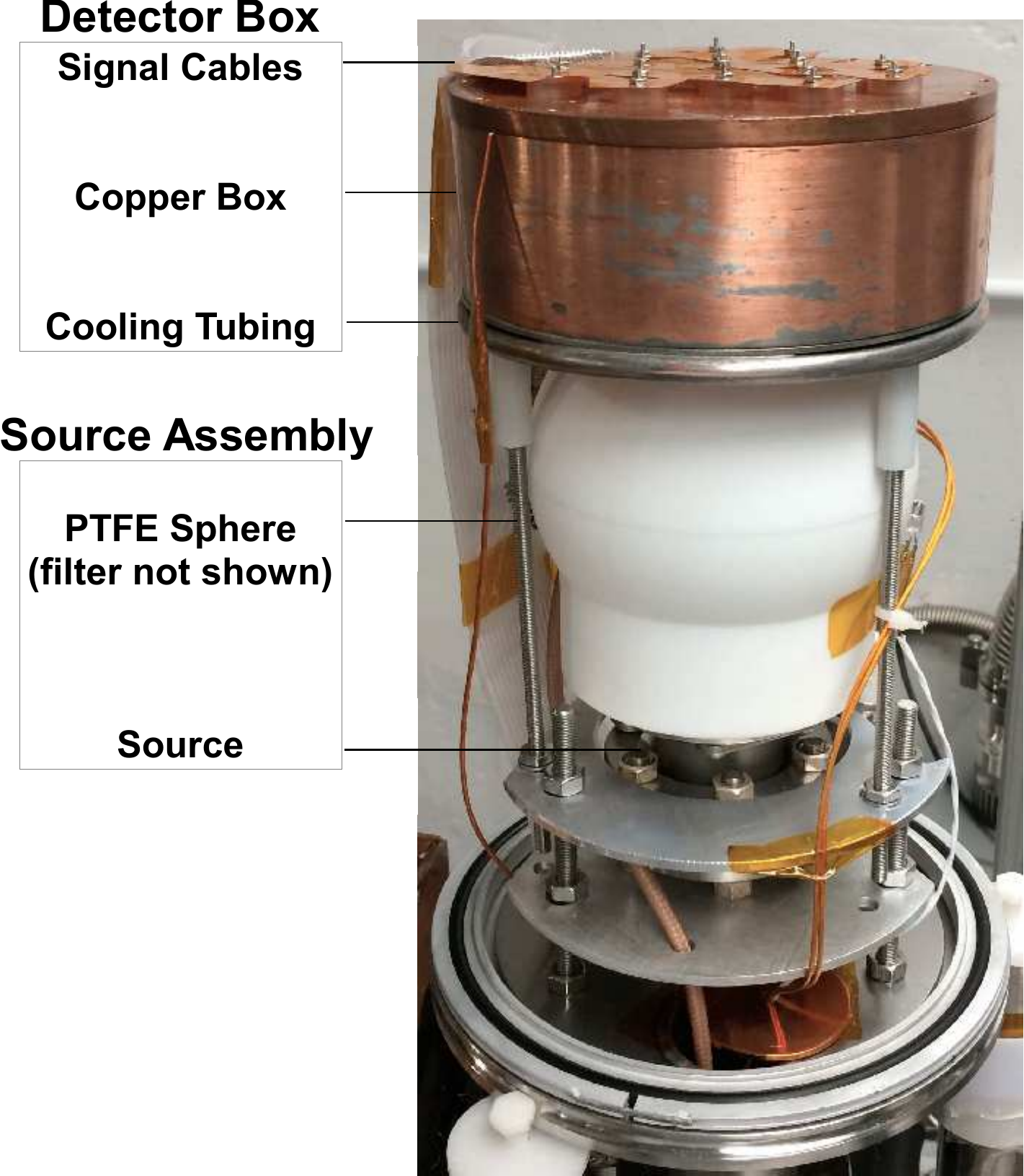}
  \caption{A cross-sectional view of the CAD model (left) and a photograph (right) of the Stanford test setup. }
 \label{fig:setup_all}
\end{figure}

\begin{figure}[htbp]
 \centering
 \includegraphics[width=0.25\textwidth]{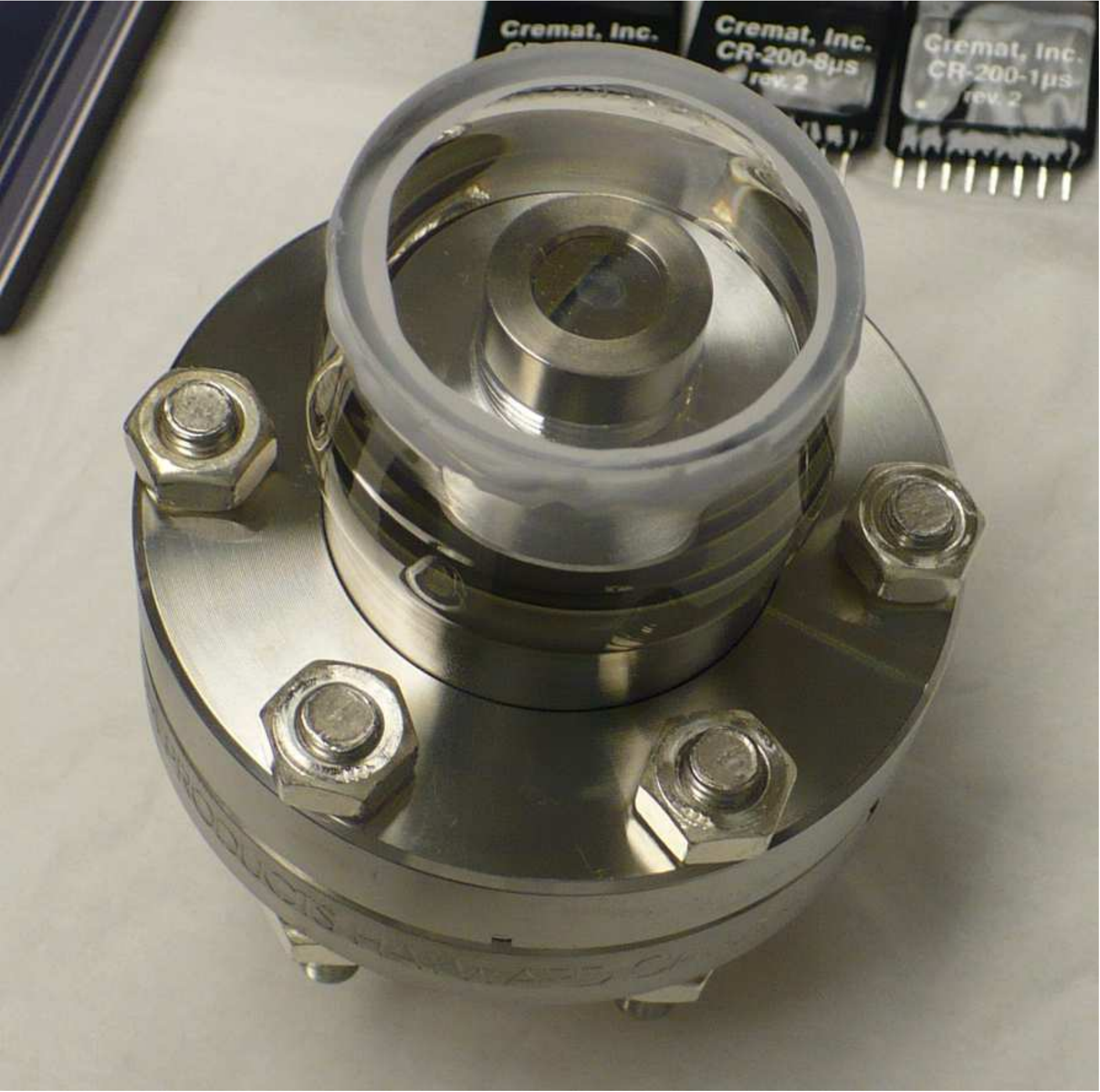}
  \caption{The custom-built xenon scintillation source.}
 \label{fig:source}
\end{figure}

The test setup is based on the earlier version employed for testing of the APDs for the EXO-200 experiment~\cite{apd:2009}. The setup was upgraded and optimized for the determination of PDE at 175-178 nm for devices of different types with surface areas up to 2$\times$2 cm$^2$. The schematic diagram and a picture of the setup are shown in Fig.~\ref{fig:setup_all}.
The main components of the set-up are the detector cage and the source assembly, both enclosed in a vacuum chamber. The detector cage is a cylindrical copper box that houses the photodetector to be tested. Different photodetectors can be mounted on the orifice (2x2 cm$^2$) in the bottom plate of the box using custom inserts that ensure equal distance from the source to the detector surface and good thermal contact between the detector and the box. The box is cooled by liquid nitrogen boil-off gas passing through a copper tube that is soldered along the circumference of the box's bottom plate. The temperature is regulated by two resistive heaters controlled with Omega~\cite{omega} PID controllers. The source assembly consists of a custom built xenon scintillation source, an opaque optical cavity (4.5 inch diameter), and a bandpass optical filter (2 inch diameter). The xenon scintillation source utilizes 83 nCi (at the moment of fabrication; roughly 60 fissions per second during measurements) of $^{252}$Cf electroplated onto a platinum surface by Eckert \& Ziegler~\cite{ez}, which is then enclosed in a miniature chamber filled with xenon gas at $\sim$1 bar pressure. A similar source based on $^{241}$Am was also fabricated, but not used routinely. 
The use of a $^{252}$Cf-fission source has the advantage of order of magnitude larger scintillation light output, compared to an alpha source (we estimate the scintillation flux at the bottom plate of the detector box to be $\sim$2 photons/mm$^2$ per fission) .
A picture of one of the two xenon scintillation sources is shown in Fig.~\ref{fig:source}. Additionally, a blue LED is installed inside the vacuum chamber and can be used as an alternative light source of varying intensity.
The optical cavity confines the xenon scintillation light emitted by the source, avoiding parasitic reflections off aluminum and Teflon surfaces inside the vacuum chamber. It is implemented as a Teflon (PTFE) sphere with two concentric openings for the source at the bottom and the detector at the top, approximately 2 inch in diameter each. Fitted at the top sphere's opening, 2-3 mm in front of the detector's surface, is the Pelham Research Optical~\cite{pelham} bandpass filter. The filter's transmission curve is centered at 180 nm (40 nm  FWHM) to avoid bias due to potential re-emission of Teflon into longer wavelengths and contamination by the sub-dominant infra-red component of xenon scintillation.

\subsection{PDE measurement}
\begin{figure}[htbp]
 \centering
 \includegraphics[width=0.35\textwidth]{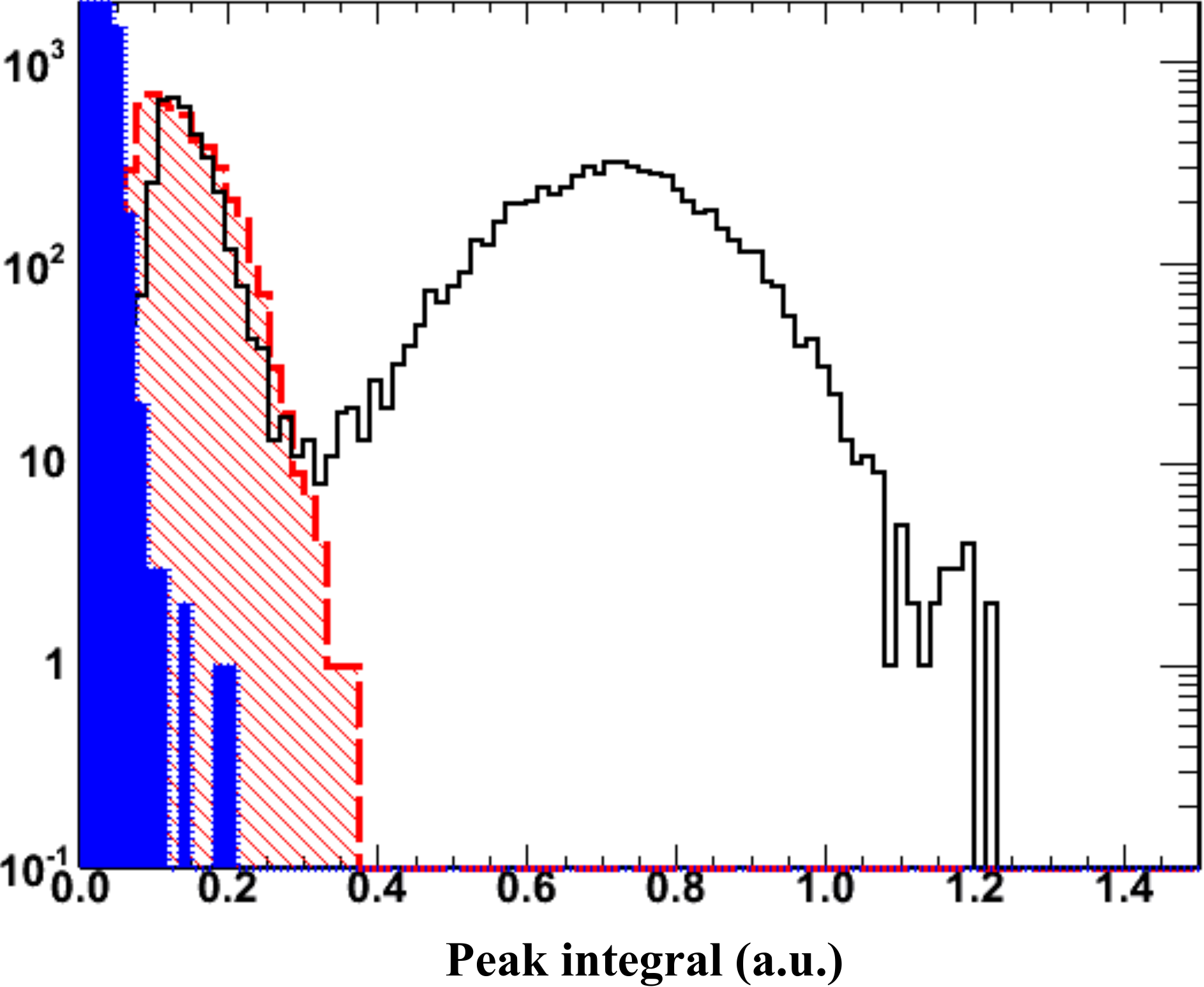}
  \caption{Charge spectrum recorded with the FBK-2010 SiPM device exposed to xenon
scintillation light excited by the $^{252}$Cf (black line) and $^{241}$Am (red hatched line) sources. The dark noise distribution is shown as blue filled region. Peaks from alpha and fission decays are visible. $^{252}$Cf undergoes both alpha and spontaneous fission decays, while $^{242}$Am only undergoes alpha decay.}
 \label{fig:peaks}
\end{figure}

Signals from the photodetectors are routed from the detector cage via a vacuum feed-through using low capacitance flexible cables and amplified by a Cremat~\cite{cremat} charge sensitive preamplifier and a Gaussian shaper. The signals are then digitized by a National Instruments digitizer. A LabView based data acquisition program collects waveforms in either continuous or triggered mode, which are then reconstructed and analyzed using ROOT~\cite{root} based custom software. Fig.~\ref{fig:peaks} shows an example spectrum obtained with the FBK-2010 SiPM device and the setup described above. 

To provide an absolute PDE reference we use a Hamamatsu R9875P photomultiplier tube (PMT) with calibrated quantum efficiency at 175 nm. The reference PMT has a synthetic silica window, an outer diameter of 15.9 mm (Cs-Te photocathode effective diameter 8 mm) and is mounted in the detector cage at the same position as other photodetectors. 
The PMT is operated at room temperature, biased to the recommended plateau voltage (1190 V), and its gain is calibrated with low intensity LED light pulses. The PDE of the SiPM devices is then extracted by comparing their responses  (specifically, the $^{252}$Cf spontaneous fission peak position, expressed in number of photoelectrons per unit area, which is typically in the range of 0.2 - 0.4 photoelectrons per mm$^2$) to that of the reference PMT.

An important part of the measurement is the parasitic charge correction applied to the SiPM devices. The total amount of observed charge is biased towards larger values, predominantly due to the contribution of correlated avalanches (typically dominated by cross-talk, with addition of after-pulses). The correction is determined from separate estimates of the rates of cross-talk and after-pulsing. Note that it is also possible for dark hits to contribute to the bias, but in practice this contribution is typically negligible.

For the PDE measurement, several sources of errors are taken into account:
\subsubsection{Raw charge spectrum calibration}

\begin{figure}[htbp]
 \centering
 \includegraphics[width=0.3\textwidth]{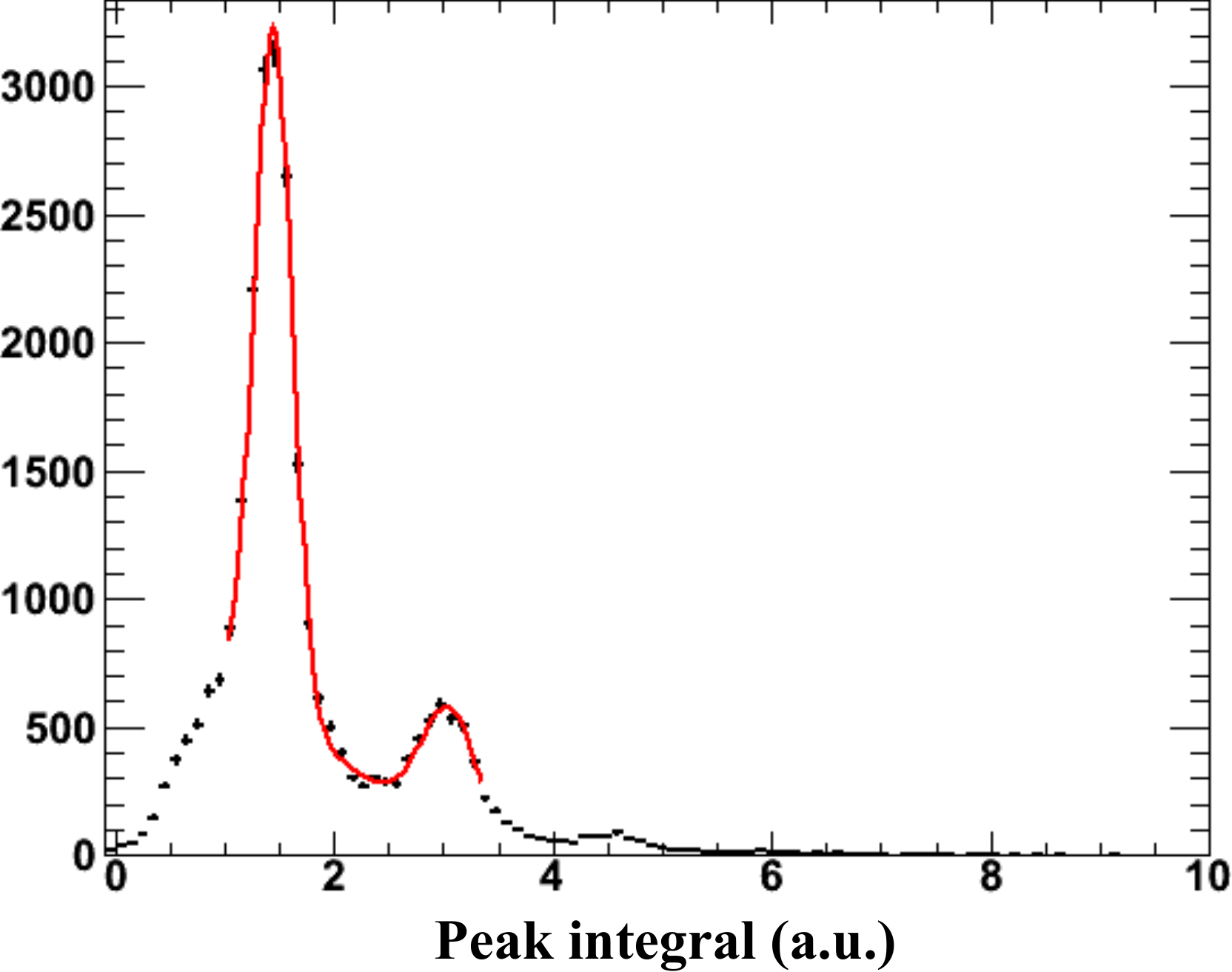}
  \caption{Dark pulse charge distribution obtained with the FBK-2010 device. Data shown as dots. The red solid line is a two Gaussian fit to the pedestal and single photoelectron peaks.}
 \label{fig:spe}
\end{figure}

\begin{figure}[htbp]
 \centering
 \includegraphics[width=0.35\textwidth]{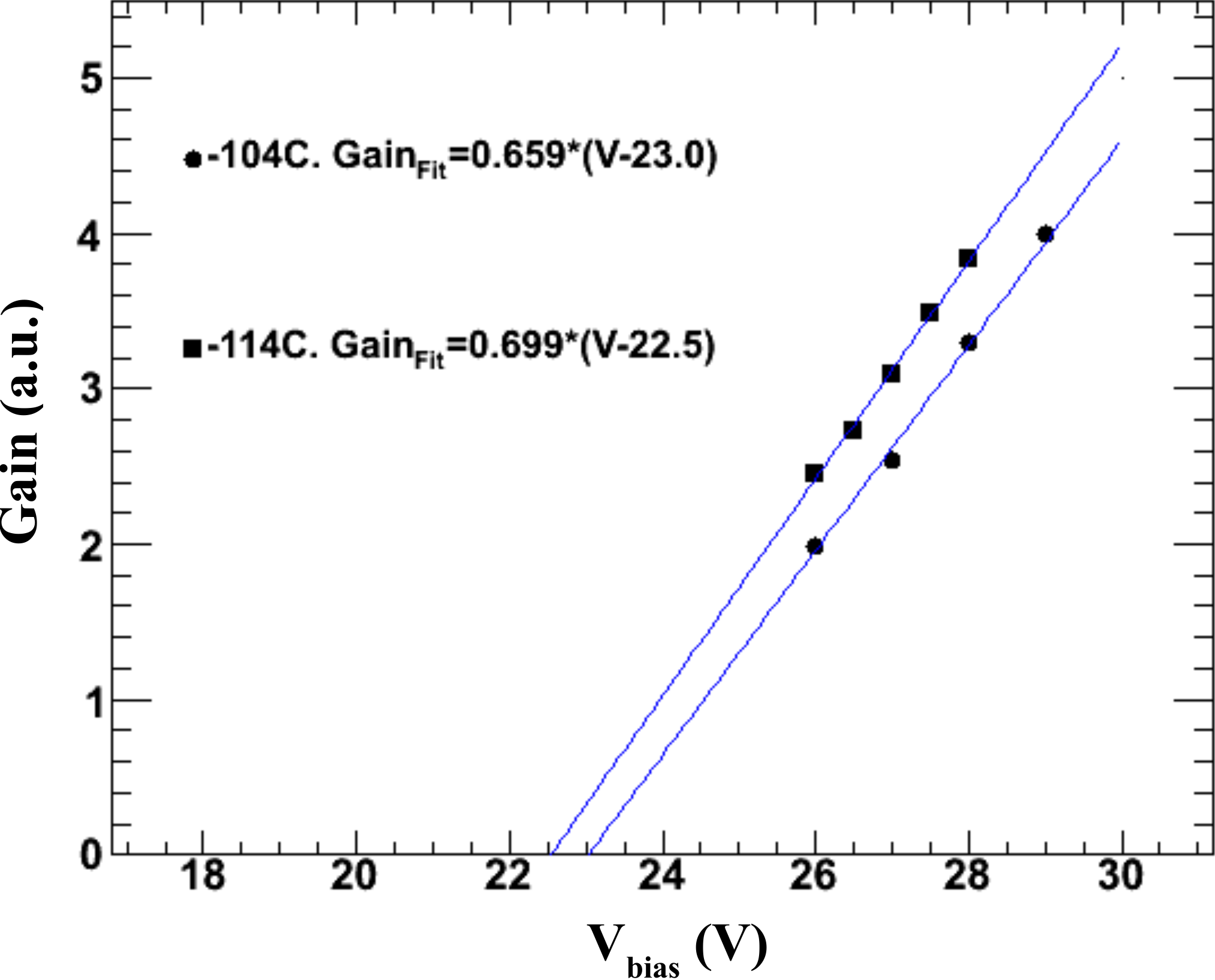}
  \caption{SiPM gain as a function of bias voltage at -104\celsius{} (circles) and -114\celsius{} (squares), measured with the FBK-2010 device. The blue lines are linear fits (the best-fit parameters are shown on the plot). Statistical error on individual gain measurements is $\sim$1\%, which is comparable to the marker size.}
 \label{fig:temp}
\end{figure}

The gain calibration is applied for each value of the over-voltage (defined as the excess bias beyond the breakdown voltage of the device) to convert the raw charge into the number of photoelectrons per unit area. The over-voltage value itself depends on the value of the breakdown voltage, which is determined by extrapolating the gain-voltage dependence to zero gain using a linear fit. The relative SiPM gain at each over-voltage is extracted from the dark charge distribution. While SiPMs are known to typically have very good single photoelectron (s.p.e.) resolution, the devices measured here have large readout areas ($\sim$1 cm$^2$, connected in parallel), which leads to increased peak widths due to a) an increased readout noise and b) an increased rate of pulse reconstruction errors related to pile-up. In practice, only single or double photoelectron peaks could be reliably resolved. The gain is then determined by fitting the dark pulse distribution with a function representing pedestal and each photoelectron peak by a Gaussian. Fig.~\ref{fig:spe} shows the distribution of random pulses occurring in the absence of light (dark pulses) obtained with the FBK-2010 device. The fit error on the s.p.e. equivalent charge is typically on the order of 0.5-1.5\%. 
This error also affects the determination of the breakdown voltage and is propagated by the linear fit of the gain-voltage dependence. Note that the breakdown voltage error (typically, $\sim$0.1-0.2 V) could be straightforwardly translated into an additional error on the gain at a given voltage, thus decoupling this uncertainty from the inherent dependence of the PDE on over-voltage. Given the linear dependence of the gain on the over-voltage, the \(\pm\)0.2 V uncertainty on the breakdown voltage translates into $\sim$4\% gain error for the FBK-2010 and $\sim$0.6\% for Hamamatsu-VUV devices, averaged over used over-voltages.
Similarly, the imperfect temperature stability of the setup (typical RMS $\sim$0.2-0.3\celsius{}) also contributes to the gain error via the over-voltage variation. To evaluate this contribution we determine the temperature dependence of the over-voltage. The FBK-2010 devices show approximately 8-10 V breakdown voltage variation between room temperature and -104\celsius{}. The rate of change of the breakdown voltage with temperature is itself a function of temperature, so we measure it close to the value of interest (Fig.~\ref{fig:temp}) and find a slope of $\sim$50 mV/\celsius{}  near -104\celsius. Conservatively assuming 1\celsius{} variation, the gain uncertainty contribution is $\sim$1\%. 
In total, a 5\% error on the charge calibration is estimated. 

\subsubsection{Solid angle}

The setup is designed such that the detectors being tested and the reference PMT are located at the same distance from the source and their surface areas subtend a small fraction of one steradian ($<$0.03 sr). In one case (Hamamatsu-VUV) a non-standard detector holder had to be used, which resulted in an offset of approximately 3 mm for this detector relative to others. This is accounted for by an appropriate correction. In all cases, the position is known to 1-2 mm that, given the distance between the source and the detectors, translates into $\approx$3\% uncertainty on the light flux. 

\subsubsection{Light source temperature dependence}

\begin{figure}[htbp]
 \centering
 \includegraphics[width=0.45\textwidth]{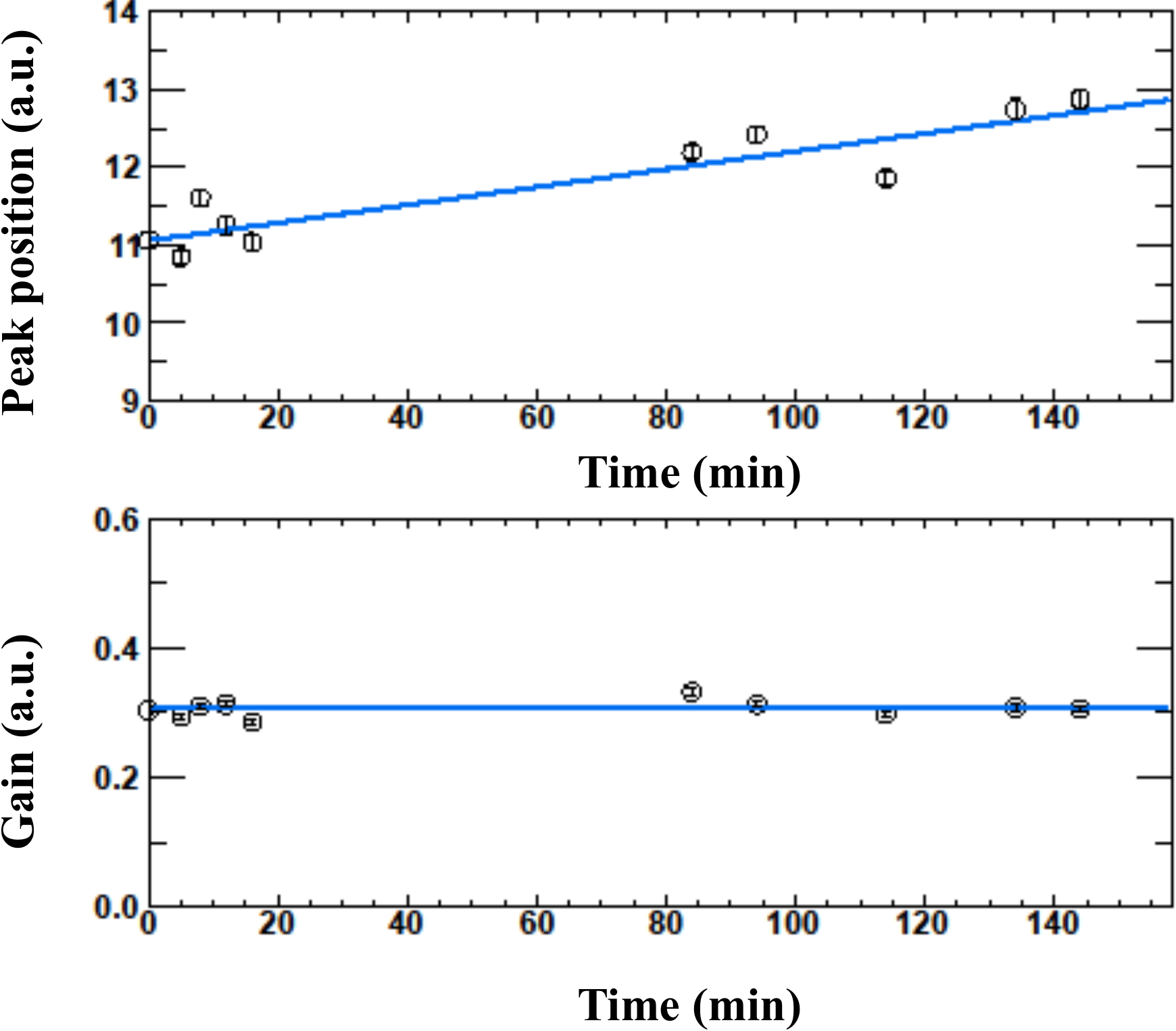}
  \caption{Drift of the fission peak position over time (top) observed with a SiPM held at -104\celsius{}. The temperature of the scintillation source was found to decrease by $\sim$4\celsius{}/hr from room temperature after the cool down of the detector cage. The SiPM gain (bottom) remained constant during the measurement, confirming that the SiPM was in thermal equilibrium with the detector cage.}
 \label{fig:tdep}
\end{figure}

Due to imperfect thermal isolation of the source assembly from the detector cage, the source temperature was observed to gradually decrease over time after the detector cage is cooled down and maintained at the liquid xenon temperature ($\sim$104\celsius{}). The source temperature decreases $\sim$4\celsius{}/hr from room temperature during the first two hours following the cool down. Coincident with the temperature decrease, the fission peak position was found to increase at the rate of $\sim$4-5\%/hr, while the SiPM gain remained constant, confirming that the SiPM remained in thermal equilibrium with the detector cage (Fig.~\ref{fig:tdep}). Conservatively assuming that measurements with different detectors are performed two hours apart after cool down, we assign a 10\% error to this effect.

\subsubsection{Fission peak equivalent charge}

The fission peak equivalent charge is determined by a Gaussian fit (Fig.~\ref{fig:meg_0}). A typical fit error is $<$1\%. While the resolution of a single device is too poor to observe any potential structures in the fission energy spectrum, the Gaussian provides a good fit, and its mean is much farther separated from the background than the alpha peak, which makes it a good variable to compare efficiency of different devices.

\subsubsection{PMT absolute reference}

\begin{figure}[htbp]
 \centering
 \includegraphics[width=0.35\textwidth]{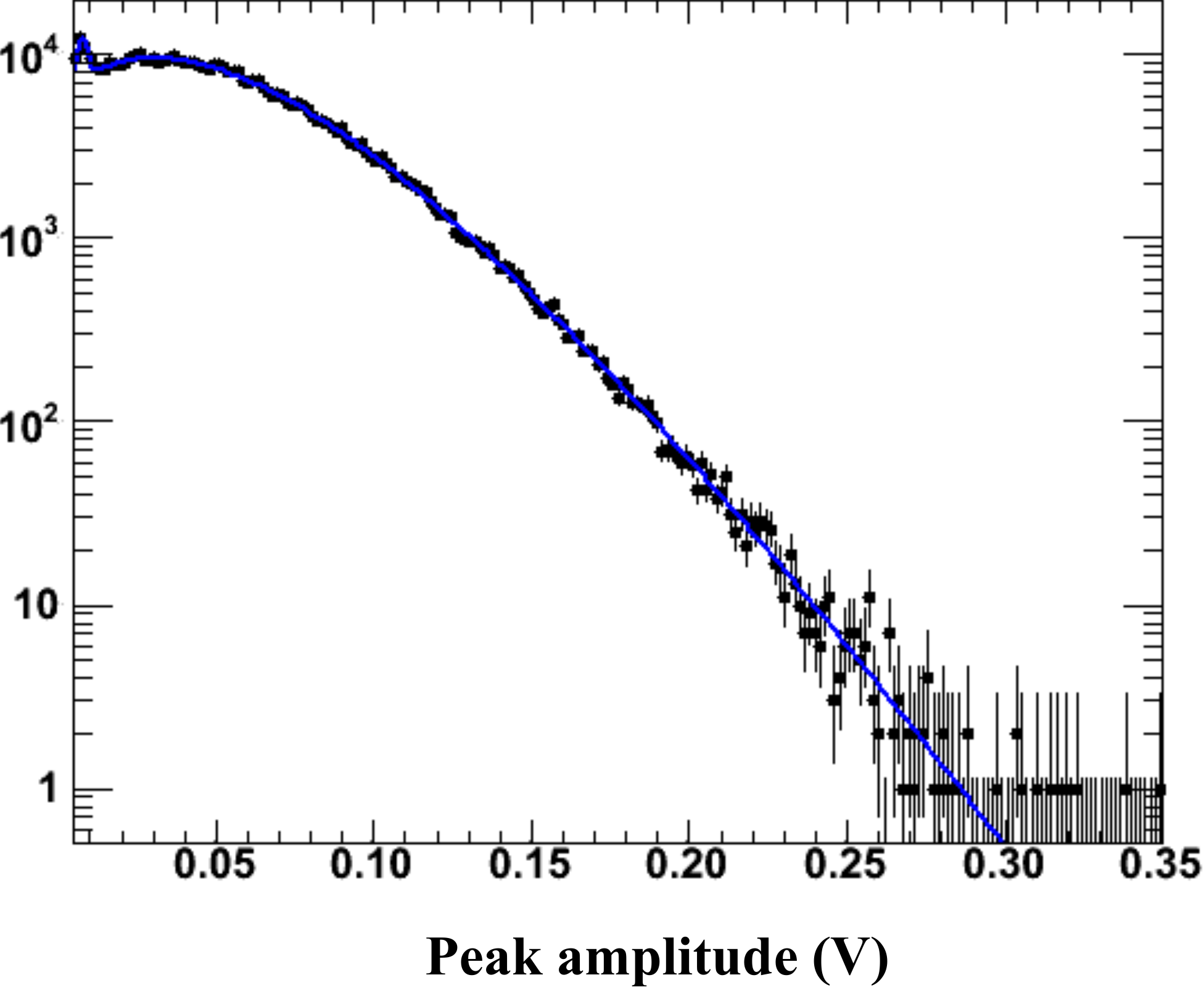}
  \caption{Distribution of peak amplitudes (black dots) obtained with the reference PMT exposed to low intensity LED light. The distribution is fit by a function (blue solid line) representing the sum of the pedestal and several photoelectron peaks. The mean Poisson rate ($\sim$2) is determined by the fraction of pedestal events.}
 \label{fig:spe_fit}
\end{figure}

The total error budget is currently dominated by the uncertainty on the absolute PDE reference, which in our case is provided by the reference PMT described earlier. The PDE of a PMT can be expressed as a product of quantum and collection efficiencies. The quantum efficiency depends on the wavelength of the detected light and was measured by Hamamatsu to be 20.1\% at 175 nm for the PMT used (the calibration uncertainty is $\sim$2\%). The collection efficiency depends on the voltage difference between the photocathode and the first dynode. It typically varies from 80\% to 100\%~\cite[Fig.4-12]{Ham:v3}, but for the PMT type used in this study was reported by Hamamatsu to be 70\%. Hamamatsu did not provide a reference curve for the particular tube we used, so we approximate the collection efficiency by 70$\pm$10\%. The central value is suggested by Hamamatsu, while the full variation from ~\cite[Fig.4-12]{Ham:v3} is used as an uncertainty. The tube to tube variability is assumed negligible. An additional uncertainty arises from imperfectly known single photoelectron gain. The PMT gain is typically calibrated with a very low intensity LED source, such that the charge spectrum obtained by triggering on the LED's driver pulse is dominated by pedestal and single photoelectrons events (the mean Poisson photoelectron rate is $\sim$0.1). This way the contamination of multiple photoelectron events is negligible and the gain can be determined typically with sub-percent uncertainty. Due to the poor peak-to-valley ratio observed for the used PMT, we calibrated the gain using larger light intensity (with mean Poisson rates of $\sim$2-3.5), such that the centroid of the resulting charge distribution is more clearly separated from the pedestal peak. A substantial remaining probability of the pedestal (i.e., zero photoelectron) events allows us to determine the mean Poisson rate, which provides the expected fractions of single-, double-, etc., photoelectron events. The fraction of the pedestal events is determined by a Gaussian fit, with the peak width and position fixed to the values obtained in the dark conditions. No loss of pedestal event is expected, given the externally triggered threshold-less data acquisition. A 10\% error on the Poisson rate is conservatively assumed. The single photoelectron gain is then determined from a multi-photoelectron fit (Fig.~\ref{fig:spe_fit}). The stability of the fit was checked by repeating the calibration procedure at different PMT voltages, recovering an exponential dependence of the gain. The gain calibration was found to vary within 6\% if the light intensity was changed such that the mean Poisson rate changed by a factor of $\sim$2.  We round the quadratic sum of the three contributions up to 19\% overall uncertainty on the PMT reference. 

\subsubsection{Parasitic charge correction}

\begin{figure}[htbp]
 \centering
 \includegraphics[width=0.35\textwidth]{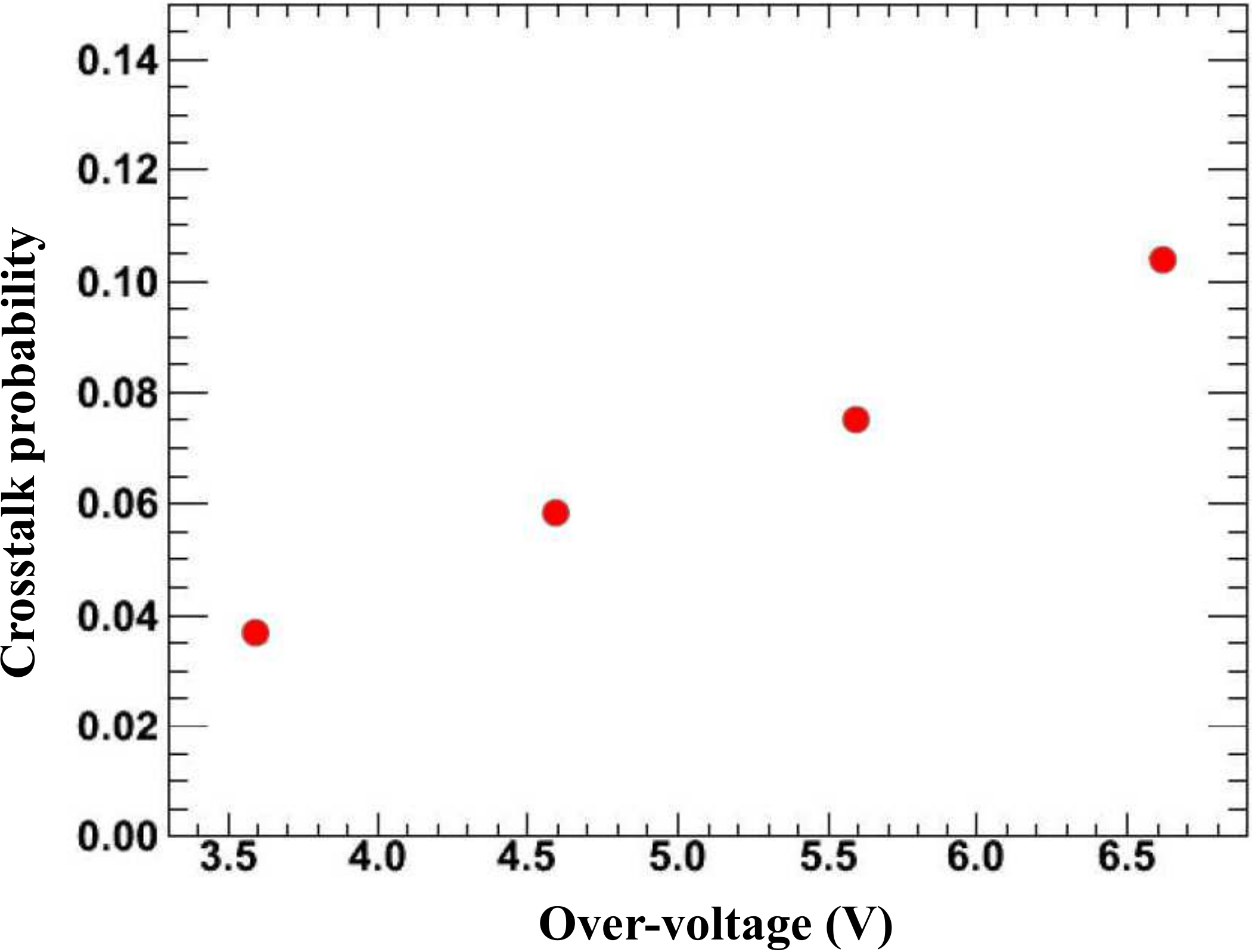}
  \caption{Probability of cross-talk, estimated from the ratio of single- to multi-photon events, as a function of over-voltage for a FBK-2010 device.}
 \label{fig:fbk_ct}
\end{figure}

Parasitic charge corrections for FBK-2010 and Hamamatsu-VUV devices are taken from published data (\cite{fbk:2013} for FBK-2010 and~\cite{MEG:2014} for Hamamatsu-VUV). In both cases, the correction takes into account the main sources of correlated noise: cross-talk and after-pulsing. The probability of either, as is pointed out in~\cite{fbk:2013}, does not change with the incoming photon flux, or wavelength. FBK claims very small probability of correlated avalanches for the devices based on their original technology, even at high over-voltages ($\sim$5\% at 5V over-voltage). For the Hamamatsu-VUV devices, on the other hand, the contribution is substantially larger ($\sim$30\% at 2V over-voltage). 
Neither publication explicitly states uncertainties on the measured correlated noise contributions to the total charge. In the case of Hamamatsu-VUV, the \(\pm\)(2-4) abs.\% spread observed in the correction values at each over-voltage for different devices is used as an estimate of the uncertainty. In the case of the FBK-2010 devices, we performed a partial cross-check by evaluating the probability of cross-talk (which is the dominant source of parasitic charge) \textit{in situ} from the ratio of single- to multi-photon signals in dark conditions. The dependence of the observed cross-talk probability on over-voltage (Fig.~\ref{fig:fbk_ct}) agrees with the parasitic charge dependence on over-voltage extracted from~\cite{fbk:2013} to within 1-2 abs.\%. Given the values of the required corrections, these differences translate into $\sim$6 and $\sim$2 rel.\% error on the PDE for Hamamatsu-VUV and FBK-2010 devices, respectively. We conservatively use the larger value for all measurements. 


Table~\ref{tab:sys} summarizes the systematic error budget. Items shown in italic correspond to errors correlated among different measured devices. The total error budget of the absolute efficiency measurement is estimated at $\sim$20\%. The relative performance of Hamamatsu-VUV and FBK-2010 SiPM devices is determined more accurately ($\sim$13\%, dominated by the light flux temperature dependence and parasitic charge correction).  

\begin{table}[t]
\caption{Systematic errors of the PDE measurement. Items shown in italic corresponds to the sources of error correlated among different measured devices.}
\centering
\begin{tabular}{l c}
\hline\hline
\textbf{Source of error} & \textbf{Value, \%} \\
\hline
\textit{PMT absolute reference} & \textit{19}\\
Light flux temperature dependence & 10 \\
Parasitic charge correction & 6\\
SiPM charge calibration & 5\\
Solid angle & 3\\
Fission peak equivalent charge & 1\\
\textbf{Total} & $\sim$\textbf{23}\\
\hline\hline
\end{tabular}
\label{tab:sys}
\end{table}

\section{Test setup for cross-talk and after-pulsing measurement}
The dark noise, after-pulsing rates, and the cross-talk probability were measured in a test setup at TRIUMF. The setup was designed to rapidly measure the performances of photodetectors in conditions as close as possible to nEXO. The photodetectors were mounted on printed circuit boards that attach to a INSTEC HCP302 cooling chuck whose temperature is regulated by the combination of liquid nitrogen boil off gas running and a heater embedded in the chuck. The cooling chuck can easily reach -110\celsius{} even when surrounded by warm gas. The whole setup is enclosed in a light tight box filled with nitrogen gas whose purpose is to prevent condensation and to allow the propagation of 175 nm light. No light sources were used, however, for the data reported in this paper.  All data reported in this paper were taken at -100\celsius{} cold chuck temperature. The photodetector temperature tracked the temperature of the chuck within a few degrees depending on the photodetector package. 

Due to several constraints, different devices were tested at TRIUMF and at Stanford (see Table~\ref{tab:devices}). The devices tested at TRIUMF were manufactured by KETEK, FBK and Hamamatsu. The KETEK devices are 3$\times$3mm$^2$, 50$\mu$m pitch SiPM. The FBK devices, FBK-2013, are 1$\times$1mm$^2$ second generation SiPMs.
The Hamamatsu devices, Hamamatsu-VIS, are 3$\times$3mm$^2$, 50$\mu$m pitch non-VUV sensitive SiPMs (also called by Hamamatsu Multi-Pixel Photon Counters, MPPCs) whose nuisance parameters are expected to be very similar to the MPPCs developed for the MEG experiment. The performances of these devices were measured at about 5V over-voltage for the KETEK and FBK-2013 devices, and 2.5V for the Hamamatsu-VIS MPPC. 

\begin{figure*}
 \centering
\includegraphics[scale=0.3]{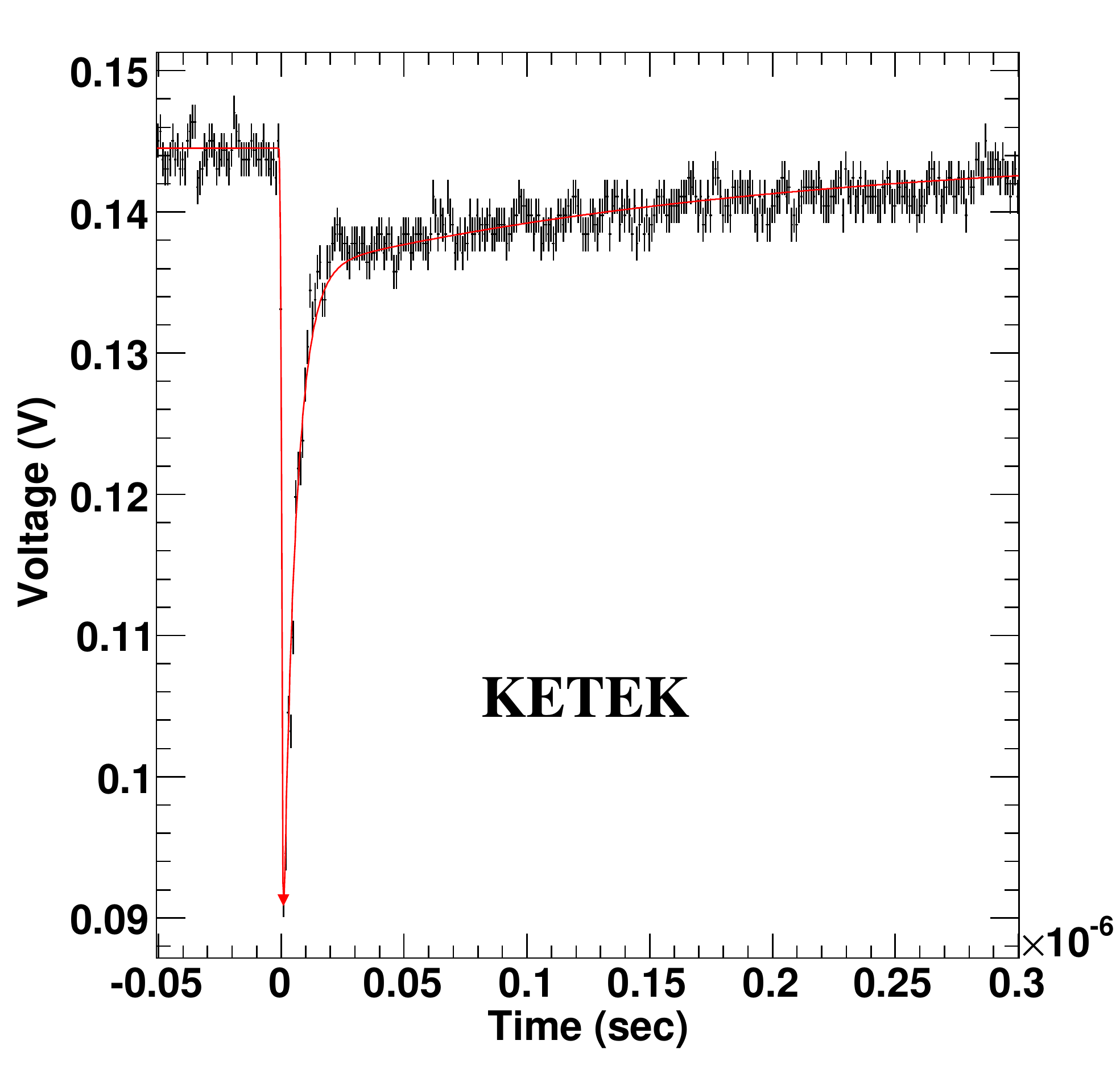}
\includegraphics[scale=0.3]{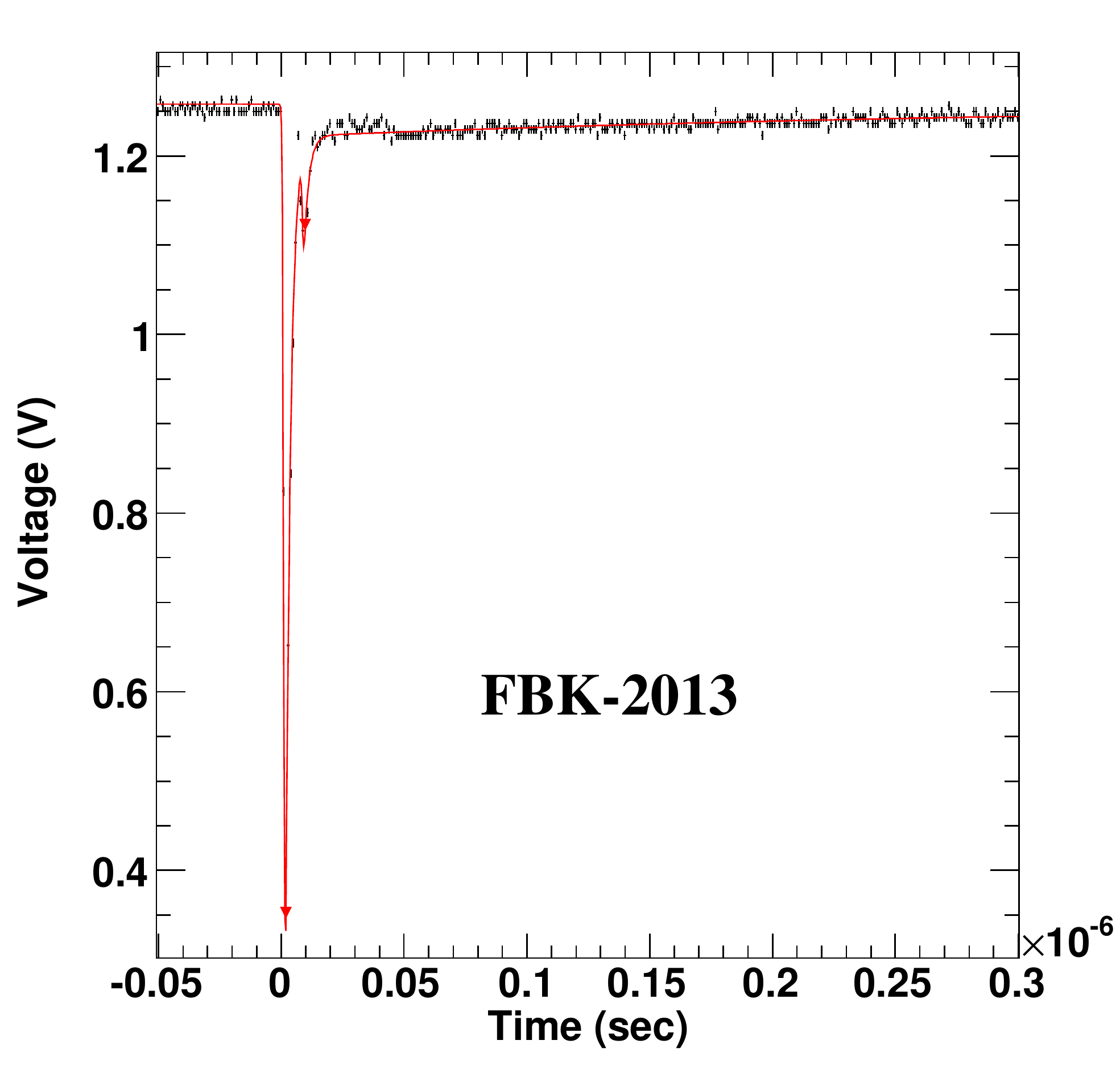}
\includegraphics[scale=0.3]{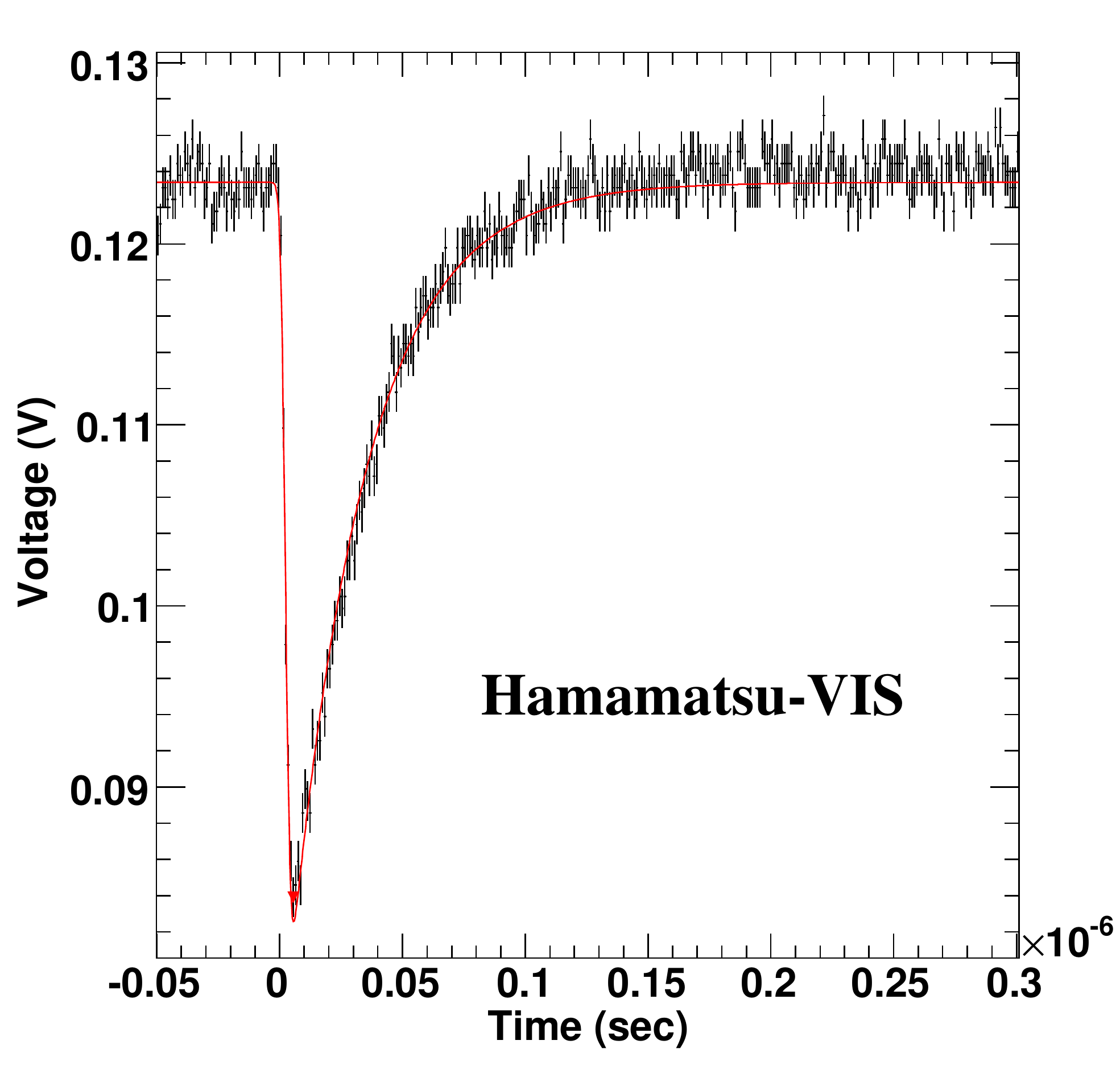}
 \caption{Examples of single avalanche pulse waveforms for a KETEK (left), FBK-2013 (middle), and Hamamatsu-VIS (right) SiPMs devices with the fit function superimposed. The function fits the prompt peak but it is truncated when displayed.}
 \label{fig:waves}
 \end{figure*}

Signals from the photodetectors were amplified by a factor 100 using two Mini-Circuits MAR6-SM+ amplifiers in series~\cite{MAR6}. A 50 Ohm coaxial cable was used to connect the photodetector to the amplifier board. The MAR6-SM+ are very fast amplifiers that allow measuring the photodetector pulse shape with minimum distortions. Signals from the amplifier board were readout by a Lecroy Waverunner oscilloscope. Waveforms were recorded by triggering on signals above the electronics noise. The waveforms were 10 $\mu$s long with the sampling rate limited to 1 Giga-sample per second (GS/s) as a compromise between timing resolution and data processing speed. Examples of digitized waveforms are shown on Fig.~\ref{fig:waves}.

Dark noise and after-pulse rates are measured by compiling the distribution of the time differences between the trigger pulse and the next pulse. At -100\celsius{} most of the waveforms have only one pulse and the timing difference is calculated using the time stamp of each waveform provided by the oscilloscope. Unfortunately, the oscilloscope has a dead time of 4 $\mu$s after each waveform that has to be accounted for when characterizing the timing distribution. After-pulses can occur within the same waveform as the trigger pulse. The waveforms are analyzed as follows: 1) a pulse finder identifies pulses whose shape and amplitude are inconsistent with electronics noise, 2) pulses are fitted using an analytical template representing the pulse shape by minimizing an approximate chi-square that is calculated using the single bin noise (about 1 mV), 3) if the chi-square is larger than a certain threshold the pulses are refitted by adding additional pulses until either the chi-square falls below the threshold or the chi-square does not improve significantly or the number of added pulses reaches a maximum of 5. The third procedure allows identifying partially overlapping pulses. The pulse template that fits all tested SiPMs is the convolution of a Gaussian distribution with one or two exponential functions.The pulse polarity is negative starting with a fast fall time and followed by a slower rise time. The fall time is calculated as the time it takes to for the pulse to go from 10\% to 90\%. The rising part of the pulse is described by exponential functions and parametrized by the exponential time constants. Two exponentials are typically needed when the parasitic capacitance across the quenching resistor is larger than a few femtofarad. 

The obtained distribution of the time differences between the trigger pulse and the next pulse is fit by a function to extract dark noise and after-pulse rates. The function is constructed assuming that both dark noise and after-pulsing affect the time difference distribution as described in~\cite{Fit:2011}. The fit function is expanded to account for several features that were either ignored or that are specific to the nEXO setup. Recovery was ignored in~\cite{Fit:2011}, but it has a noticeable impact on the probability of generating and detecting after-pulsing. Indeed, the probability of triggering an avalanche of any kind depends on the over-voltage and our fit function assumes somewhat arbitrarily that it scales linearly with over-voltage, hence it follows the exponential recovery function. In the early part of the recovery, our pulse finder algorithm may miss pulses if their amplitudes are too small. We model this process by introducing a dead time that depends on the recovery time constant (150, 100, and 8 ns for the KETEK, FBK-2013 and Hamamatsu-VIS SiPMs, respectively). Our function ignores delayed cross-talk that do not suffer from recovery because their occurrence rate is relatively small. They could be included at the cost of yet another parameter accounting for delayed cross-talk probability.  Our function also includes the oscilloscope deadtime between 8.5 and 12.5 $\mu$s. The complete function shown in~\eqref{eq:DTimeComb} combines the functions shown in~\eqref{eq:AP},~\eqref{eq:NOAP},~\eqref{eq:DN}, and~\eqref{eq:NODN}. The dark noise rate is handled by a single rate parameter, $R_{D}$. Each after-pulsing probability is parameterized by an exponential function with two free parameters - a time constant, $\tau_{A_{i}}$, and a probability,  $P_{A_{i}}$. The after-pulsing indexes are omitted in~\eqref{eq:AP} and~\eqref{eq:NOAP} for simplicity. After-pulsing is approximated using a probability rather than using Poisson statistics for simplicity as well. This approximation is justified as long as all after-pulsing probabilities are low. $\tau_{r}$ is the recovery time constant, $t_{rd}$, $t_{sds}$, and $t_{sde}$ are the recovery deadtime, oscilloscope start deadtime, and oscilloscope end deadtime respectively.

\small
\begin{fleqn}[0pt]
\begin{align}
P_{total}(t) &= P_{D}(t)\prod^{i<n_{A}}_{i=0}P_{\bar{A}_i}(t)\nonumber\\
& + P_{\bar{D}}(t)\sum^{i<n_{A}}_{i=0}[P_{A_i}(t)\prod^{j<n_{A} \: \& \: j\ne i}_{j=0}P_{\bar{A}_j}(t)]
\label{eq:DTimeComb}
\end{align}

\begin{equation}
P_{A}(t) =  
\begin{cases}
  0, &\text{if } t {<} t_{rd} \textrm{ or } t_{sds} {<} t {<} t_{sde} \\
\frac{1-e^{-\frac{t}{\tau_{r}}}}{\tau_{A}} e^{-\frac{t}{\tau_{A}}}, &\text{otherwise}
\label{eq:AP}
\end{cases}
\end{equation}

\begin{align}
P_{\bar{A}}(t) = 
\begin{cases}
0, \hspace{4.83cm}\text{if } t {\le} t_{rd} \\
e^{-\frac{t_{rd}}{\tau_{A}}}{+}e^{-\frac{t_{rd}}{\lambda}}{-}e^{-\frac{t}{\tau_{A}}}{-}e^{-\frac{t}{\lambda}}, \hspace{1.195cm}\text{if } t_{rd} {<} t {\le} t_{sds} \\
e^{-\frac{t_{rd}}{\tau_{A}}}{+}e^{-\frac{t_{rd}}{\lambda}}{-}e^{-\frac{t_{sds}}{\tau_{AP}}}{-}e^{-\frac{t_{sds}}{\lambda}}, \hspace{0.67cm}\text{if }t_{sds} {<} t {\le} t_{sde} \\
e^{-\frac{t_{rd}}{\tau_{A}}}{+}e^{-\frac{t_{rd}}{\lambda}}{-}e^{-\frac{t_{sds}}{\tau_{A}}}{-}e^{-\frac{t_{sds}}{\lambda}}{+} \\
\quad\quad\quad e^{-\frac{t_{sde}}{\tau_{A}}}{+}e^{-\frac{t_{sde}}{\lambda}}{-}e^{-\frac{t}{\tau_{A}}}{-}e^{-\frac{t}{\lambda}}, \text{if } t {>} t_{sde} \\
\end{cases}
 \label{eq:NOAP}
\end{align}
where \(\lambda = (\tau_{r} + \tau_{A})/\tau_{r}/\tau_{A}\).

\begin{eqnarray}
P_{D}(t) = \left\{
  \begin{array}{l l}
    R_{D} e^{-R_{D} t}, & \text{if } t {\le} t_{sds}\\
    0, & \text{if  } t_{sds} {<} t {\le} t_{sde} \\
    R_{D} e^{-R_{D} (t-t_{sde}+t_{sds})}, &  \text{if } t {>} t_{sde}\\
  \end{array} \right.
\label{eq:DN}
\end{eqnarray}

\begin{eqnarray}
P_{\bar{D}}(t) = \left\{
\begin{array}{l l}
 e^{-R_{D} t}, & \text{if } t {\le} t_{sds}\\
 e^{-R_{D} t_{sds}}, & \text{if }t_{sds} {<} t {\le} t_{sde} \\
 e^{-R_{D} (t-t_{sde}+t_{sds})}, & \text{if } t {>} t_{sde} \\
\end{array}  \right.
\label{eq:NODN}
\end{eqnarray}
\end{fleqn}
\normalsize

The cross-talk probability was inferred by measuring the fraction of single pixel avalanches in the trigger window. We define the cross-talk probability as the probability that a single pixel avalanche triggers one or more additional avalanches within 1 ns of the parent avalanche. The physical mechanism responsible for cross-talk is the absorption of photons produced in the parent avalanche within the high field region of neighboring pixels. The same mechanism is known to lead to delayed avalanches, within a few ns up to several $\mu$s later due to the diffusion of the charge carriers produced by photon absorption in regions with zero or very small electric field. By construction, we only include prompt cross-talk and such a delayed cross-talk contributes to after-pulsing instead. Indeed, to measure cross-talk we calculate the probability $p_1$ that the trigger pulse is a single pixel avalanche and we define the cross-talk probability as $1-p_1$.

\section{Radio-purity measurements}
\begin{figure}[htbp]
 \centering
 \includegraphics[width=0.45\textwidth]{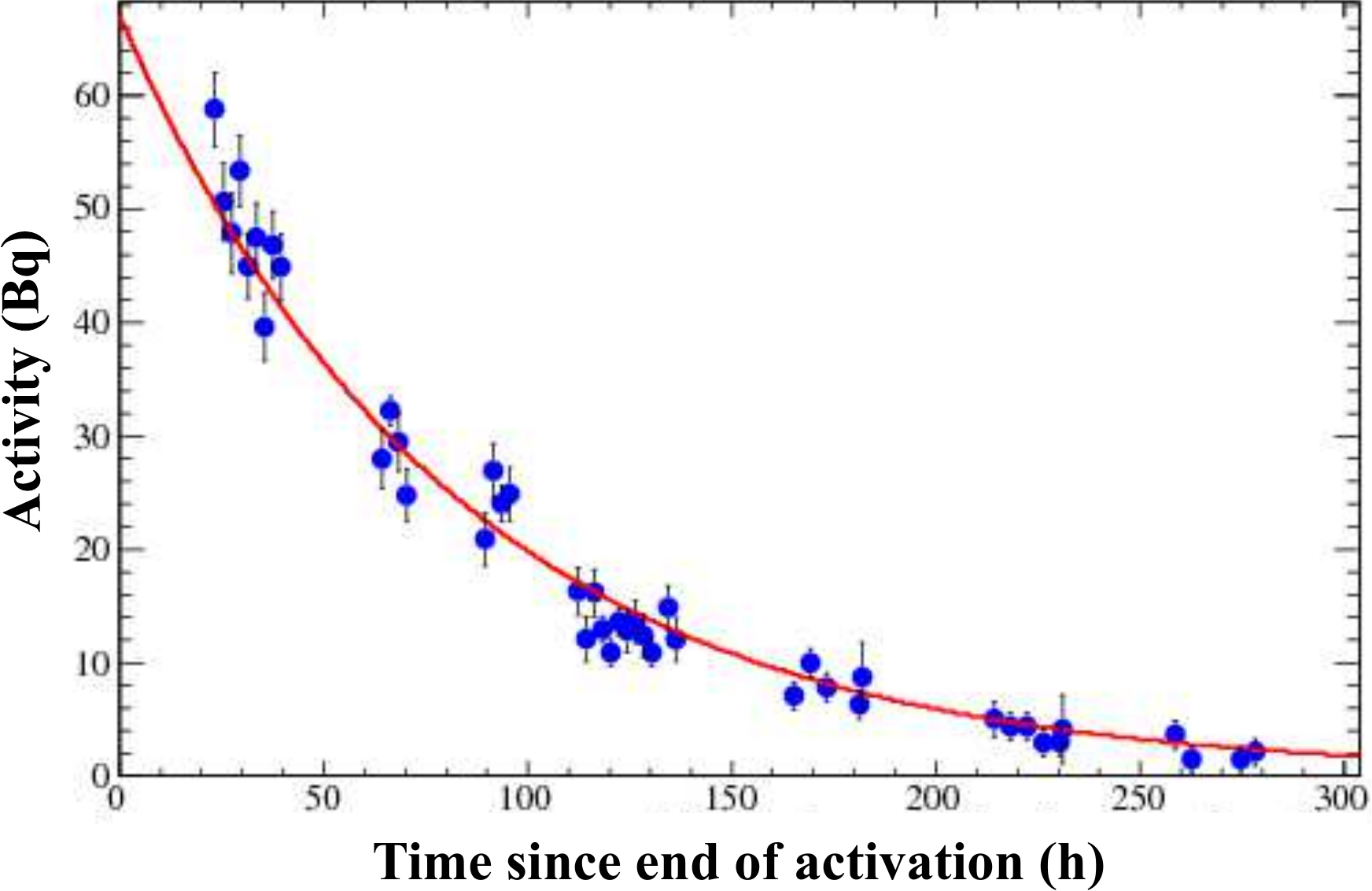}
  \caption{An example for the time dependence fit for $^{239}$Np, the blue dots are data and the red line is the best fit value.  The half-life is fixed to its tabulated value (56.54 h), so only the activity at the end of the activation (67+/-1 Bq) is determined by the fit.}
 \label{fig:Np_fixed}
\end{figure}

The nEXO group at the University of Alabama investigated the radioactivity content of FBK-2013 SiPMs by means of neutron
activation analysis (NAA). Particular attention was paid to K, U and Th. The radio active decay series of $^{232}$Th and $^{238}$U are a background concern for nEXO. High energy gamma-rays, emitted during the decay of Th and U daughters, can penetrate the detector’s interior and contribute to the measured background. Potassium has a lower Q-value, so is not important for the neutrinoless DBD, but it can affect the accuracy of alternative physics searches. Table~\ref{tab:Qvalue} shows the half-lives and  $\gamma$-ray energies for the important nEXO backgrounds. Direct gamma-ray counting of large samples allows us to measure Th, and U concentrations of a few hundred ppt. NAA allows for ppt and even sub-ppt sensitivities and requires smaller samples, than direct background counting, an important aspect for this study where kg-size samples are hard to obtain. However, the translation of NAA-determined Th and U concentrations into a background
expectation value requires assumptions on the chain equilibrium, not required for direct background counting.
 \begin{table}
    \caption{Parameters of the important $\gamma$ backgrounds for nEXO~\cite{nudat}.}
     \centering
     \begin{tabular}{ l l l l}
       \hline
       \hline
       Parent Isotope	& half-life 			&Important decay	&Important $\gamma$-ray				\\ 
       				&[1$\cdot$10$^{9}y]$&				& [keV]		\\\hline
       $^{40}$K		&1.248$\pm$ 0.003	&$^{40}$K $\rightarrow ^{40}$Ar & 1460.822$\pm$0.006			\\
       $^{232}$Th	&14.0  $\pm$ 0.1		&$ ^{208}$Tl$\rightarrow ^{208}$Pb&  2614.511 $\pm$ 0.10	\\	
       $^{238}$U		&4.468 $\pm$ 0.003	&$^{214}$Bi$\rightarrow ^{214}$Po&2447.70 $\pm$ 0.3\\ \hline \hline
     \end{tabular}
     \label{tab:Qvalue}
   \end{table}
Three samples obtained from FBK were activated at the MIT research reactor (MITR) in three separate beam times. In order to evaluate bulk and not surface properties of these devices, they were cleaned using the techniques described in~\cite{Leonard:2008}. However,
there were some variations from the technique described in~\cite{Leonard:2008} and what was done here. The June 2013 activation sample was soaked for three days in 0.5M certified low Th/U nitric acid~\cite{acid}. During the preparation process the TiN backing to the FBK-2013 chips was lost. To counter the loss of the backing the acid soak was reduced to 12 h for the following activations. 
The integral thermal, epi-thermal, and fast fission neutron fluxes at MITR were determined by activating NIST certified fly ash~\cite{NIST:2004} (an element cocktail
 with known composition) and a sample of TiN. In earlier studies we established that sample port 2PH1 (located close to the fuel element) has a high thermal neutron flux of about (4-5)$\cdot$10$^{13}$ cm$^{-2}$s$^{-1}$ but also a relatively high fast neutron flux of about 3$\cdot$10$^{12}$ cm$^{-2}$s$^{-1}$. This leads to the creation of significant Ti-related side activities that interfere with the Th and U analysis. These unwanted side activities are mainly due to the $^{46}$Ti(n,p)$^{46}$Sc and $^{48}$Ti(n,p)$^{48}$Sc reactions.
To suppress these backgrounds we activated the last two SiPM samples in MITR’s sample port 1PH1, showing a 230-fold reduced fast neutron flux at the expense of a 6-times lower thermal neutron flux. We compensated for the lower thermal flux by means of longer exposure times, a viable solution given the relatively long live time of the Th and U activation products $^{233}$Pa and $^{239}$Np. For 1PH1, the fast neutron flux was found to be so low,
that the fly-ash could not provide a definitive measurement. To find the fast-neutron flux a sample of TiN was also activated.
After receiving the activated samples at the University of Alabama they were separated from the activation vials and then weighed to account for sample loss. Time and energy differential counting was then performed using two shielded Ge-detectors. The calibration of the gamma ray detection efficiency of the Ge detectors was performed using an Eckert \& Ziegler~\cite{ez} source solution, containing multiple gamma-ray emitters of known activity. The gamma-rays counted by the detector, in form of full absorptions peaks, were compared to the known
activities to arrive at an energy dependent efficiency curve, described by a fifth degree polynomial fit. For the analysis of the samples, the code described in~\cite{Leonard:2008} was used. Briefly, this allows the use of the full data set and produces a time dependent decay curve for each isotope. An example of the double differential energy and time analysis is shown in Fig.~\ref{fig:Np_fixed}. The fitted sample activity at some reference time (N$_0$ in Fig.~\ref{fig:Np_fixed}) then summarizes all $\gamma$-peak  integrals evaluated at all times in form of a single value. For all sample fits the half lives are fixed to their tabulated values. The elemental concentrations were then inferred using properly averaged thermal/epi-thermal and fast neutron cross-sections~\cite{JENDL:2011}. It was assumed that the relative isotopic abundances in the samples are given by the natural  terrestrial abundance ratios. Table~\ref{tab:samples} shows the details of the three beam times performed with FBK-2013 chips. For the first two activation campaigns, sensitive analyses of K and U were hampered by the relatively long
delay between end of irradiation and start of counting.

   \begin{table}[t]
    \caption{Listing of the samples analyzed in three activation campaigns at MITR.
           We list the samples masses, MITR sample port,
           and counting delay (after end of activation). The May 2014 activation was split into two time segments as the sample was ejected early. This is taken into account in the analysis.}
     \centering
     \begin{tabular}{ l l l l l}
       \hline
       \hline
       Activation	&Port	&Sample			&Activation	&delay in 	\\ 
       			&		& mass [g]		&length [h]	&counting [h]\\\hline
       June 2013	&2PH1	&0.1705			&8.00		&103\\
       May 2014	&1PH1	&3.5519			&81.43	 &127\\	
       July 2014	&1PH1	&3.8906			&80.45		&33\\ \hline \hline
     \end{tabular}
     \label{tab:samples}
   \end{table}

\section{Results}

\subsection{Photon detection efficiency at 175-178 nm}

\begin{figure}[htbp]
 \centering
 \includegraphics[width=0.35\textwidth]{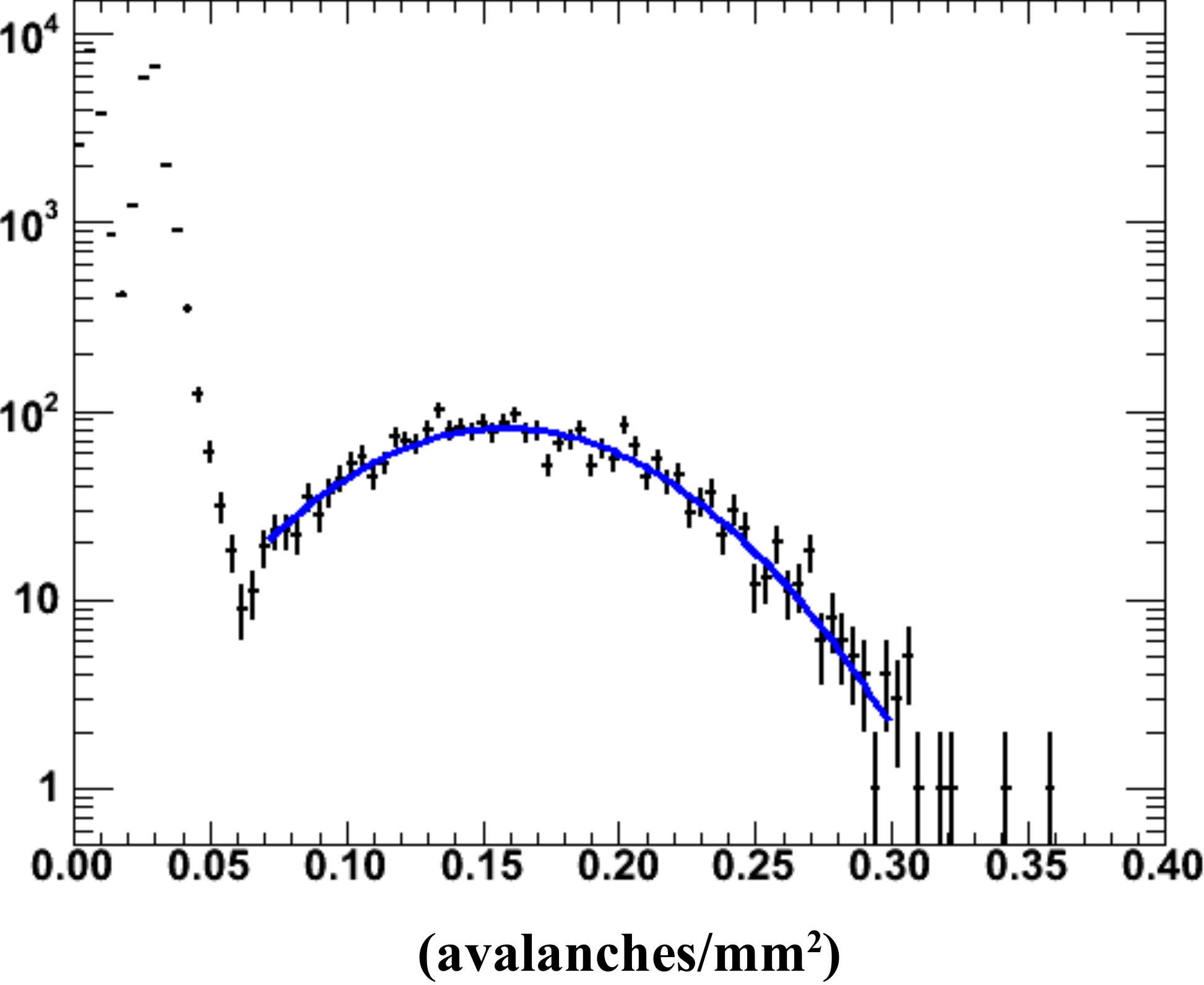}
  \caption{Charge spectrum from a Hamamatsu-VUV device obtained at 1.5 V over-voltage. The horizontal axis represents the number of avalanches (true photoelectrons plus correlated cell discharges) per mm$^2$. The data are shown as black dots. The blue solid line depicts a Gaussian fit to the fission fragment peak.}
 \label{fig:meg_0}
\end{figure}

Fig.~\ref{fig:meg_0} shows an example of the xenon scintillation spectrum obtained with the Hamamatsu-VUV device in the Stanford setup. The fission fragment peak equivalent charge is determined with a Gaussian fit. The correction for parasitic charge is applied based on~\cite{MEG:2014}. 

Two additional corrections are needed in this particular case. First, the device's packaging includes a 0.5 mm thick quartz window in front of the sensitive area. To compensate for the finite transmittance of the window, we apply a 10$\pm$1\% correction~\cite[Fig.5]{ootani:2014}. Second, due to specifics of the device's packaging, the sensitive surface was positioned 3$\pm$1 mm farther away from the light source than other detectors and the reference PMT. This results in 5.0$\pm$1.5\% correction for the smaller solid angle. 

Table~\ref{tab:pde} lists fission peak positions, correction values, and PDE numbers at different over-voltages for the measured devices. 
All measurements were performed at -104\celsius{}.  

\begin{figure}[htbp]
 \centering
 \includegraphics[width=0.45\textwidth]{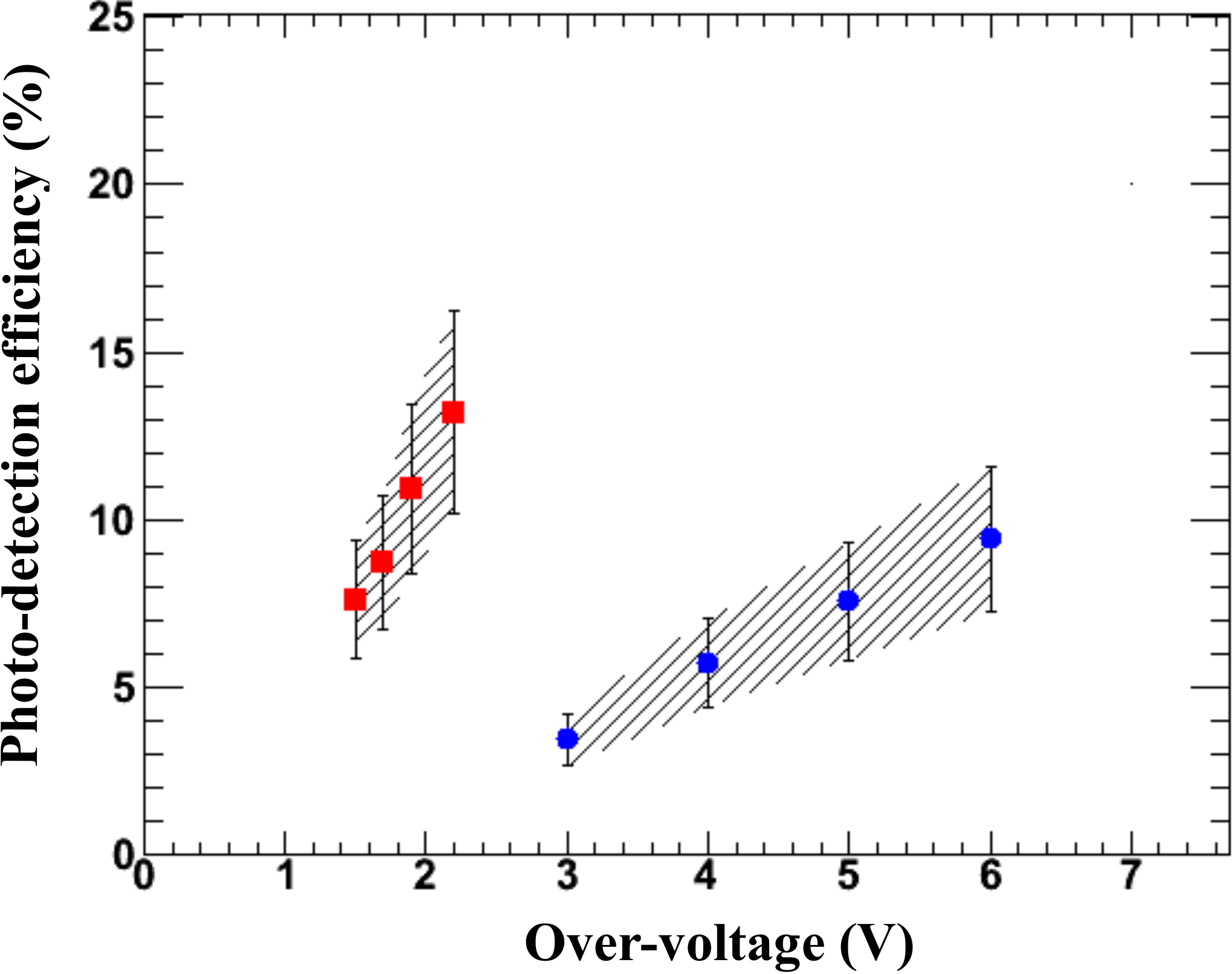}
  \caption{PDE at 175-178 nm as a function of over-voltage for Hamamatsu-VUV (red squares) and FBK-2010 (blue dots) devices. Hatched areas represent the error bands.}
 \label{fig:pde}
\end{figure}

\begin{table}[t]
\caption{PDE results. Note an additional correction of 15\% has been added in the case of Hamamatsu-VUV devices, as described in the text. Relative uncertainty $\sim$23\%.}
\centering
\begin{tabular}{l c c c c }
\hline\hline
\parbox[t]{0.5cm}{Device} &\parbox[t]{1.75cm}{\textbf{Over-voltage, [V]}} & \parbox[t]{1.3cm}{\textbf{Fission peak, [p.e./mm$^2$]}} & \parbox[t]{1.3cm}{\textbf{Parasitic charge correction, [\%]}} &\parbox[t]{0.5cm}{\textbf{PDE, [\%]}} \\
\hline
\textit{Hamamatsu-VUV}& 1.5 & 0.135 & 14.4 & 7.6\\
			& 1.7 & 0.161 & 19.2 & 8.7\\
			& 1.9 & 0.174 & 24.0 & 11.0 \\
			& 2.2 & 0.269 & 31.3 & 13.2 \\
\textit{FBK-2010}		& 3 & 0.064 & 2.2 & 3.4 \\
			& 4 & 0.110 & 4.4 & 5.7\\
			& 5 & 0.147 & 6.7 &  7.6 \\
			& 6 & 0.187 & 8.9 & 9.4 \\
\hline\hline
\end{tabular}
\label{tab:pde}
\end{table}
Fig.~\ref{fig:pde} shows the PDE as a function of over-voltage for the Hamamatsu-VUV and FBK-2010 devices. Note that while at similar over-voltages the FBK-2010 devices show several times lower PDE than Hamamatsu-VUV's, they can tolerate much larger over-voltages due to substantially smaller rates of correlated avalanches, eventually reaching comparable PDE values.

\subsection{Photodetector nuisance parameters and timing analysis}
Fig.~\ref{fig:waves} shows typical waveforms for the KETEK, FBK-2013, and Hamamatsu-VIS SiPMs along with the fitted function. 
The parameters of the function that best reproduce each pulse shape are summarized in Table~\ref{tab:nuisanceSum}. 
Hamamatsu-VIS MPPC exhibits a single rise time constant and a relatively slow fall time compare to the KETEK and FBK-2013 SiPMs. 
The KETEK and FBK-2013 SiPMs have similar pulse shapes with a very sharp peak first followed by a very slow decay time constant.
Most of the charges produced in the avalanches is contained in the later part of the pulse shape.

\begin{table*}[t]
\caption{Operational, nuisance, and pulse shape parameters of the SiPMs tested at -104\celsius.}
\centering
\begin{tabular}{l c c c}
\hline\hline
&Ketek & FBK-2013 & Hamamatsu-VIS \\ 
\hline
Operating Voltage (V) & 29.50 & 28.50 & 59 \\
Breakdown Voltage (V) & 24.70 $\pm$ 0.05 & 23.50 $\pm$ 0.05 & 56.43 $\pm$ 0.05 \\
Pulse fall time (10 to 90\%) (ns) & 0.9 & 1.0 & 3.5 \\
Pulse rise (exponential) time constant (ns) & 5.3 & 2 & 30.6 \\
Pulse 2nd fall time constant (ns) & 200 & 300 & N/A \\
Pulse 2nd fall time fraction & 0.14 & 0.023 & N/A \\
Crosstalk probability &0.253 $\pm$0.001 & 0.274 $\pm$ 0.001 & 0.249 $\pm$ 0.002 \\
Recovery time, $\tau_{r}$, (ns) & 750 $\pm$ 25 & 360 $\pm$ 10 & 21 $\pm$ 1 \\
Dark noise rate, $R_{D}$, (Hz/mm$^{2}$) & 8.0 $\pm$ 0.2 & 310 $\pm$ 1 & 0.50 $\pm$ 0.01 \\
Total after-pulsing probability      & 0.57 $\pm$ 0.02  & 0.025 $\pm$ 0.001 & 0.16 $\pm$ 0.01 \\
After-pulsing within 1 $\mu$s & 0.18 $\pm$ 0.02  & 0.016 $\pm$ 0.001 & 0.033 $\pm$ 0.005 \\
1$^{st}$ After-pulsing time constant, $\tau_{A_1}$, (ns) / probability, $ P_{A_1}$     & 235 / 0.18 & 179 / 0.01  & 8.3 / 0.031\\
2$^{nd}$ After-pulsing time constant, $\tau_{A_2}$, ($\mu$s) / probability, $ P_{A_2}$ & 2.5 / 0.04 & 0.87 / 0.01  & 8   / 0.01 \\
3$^{rd}$ After-pulsing time constant, $\tau_{A_3}$, ($\mu$s) / probability, $ P_{A_3}$ & 160 / 0.35 & 8.2 / 0.006 & 200 / 0.05 \\
4$^{th}$ After-pulsing time constant, $\tau_{A_4}$, ($\mu$s) / probability, $ P_{A_4}$ & N/A / 0    & N/A / 0     & 5000 / 0.07 \\
\hline\hline
\end{tabular} 
 \label{tab:nuisanceSum}
\end{table*}

The cross-talk probabilities measured for all 3 devices are shown in Table~\ref{tab:nuisanceSum}. The probabilities are remarkably similar for all 3 devices. They unfortunately all exceed 20$\%$, which is our upper limit for the correlated avalanche probability that includes cross-talk and after-pulsing. The systematic errors on this measurement are small (on the order of a few percent) as it only requires proper identification of the single avalanche pulse which is very well defined for all devices

  \begin{figure}
 \centering
\includegraphics[scale=0.45]{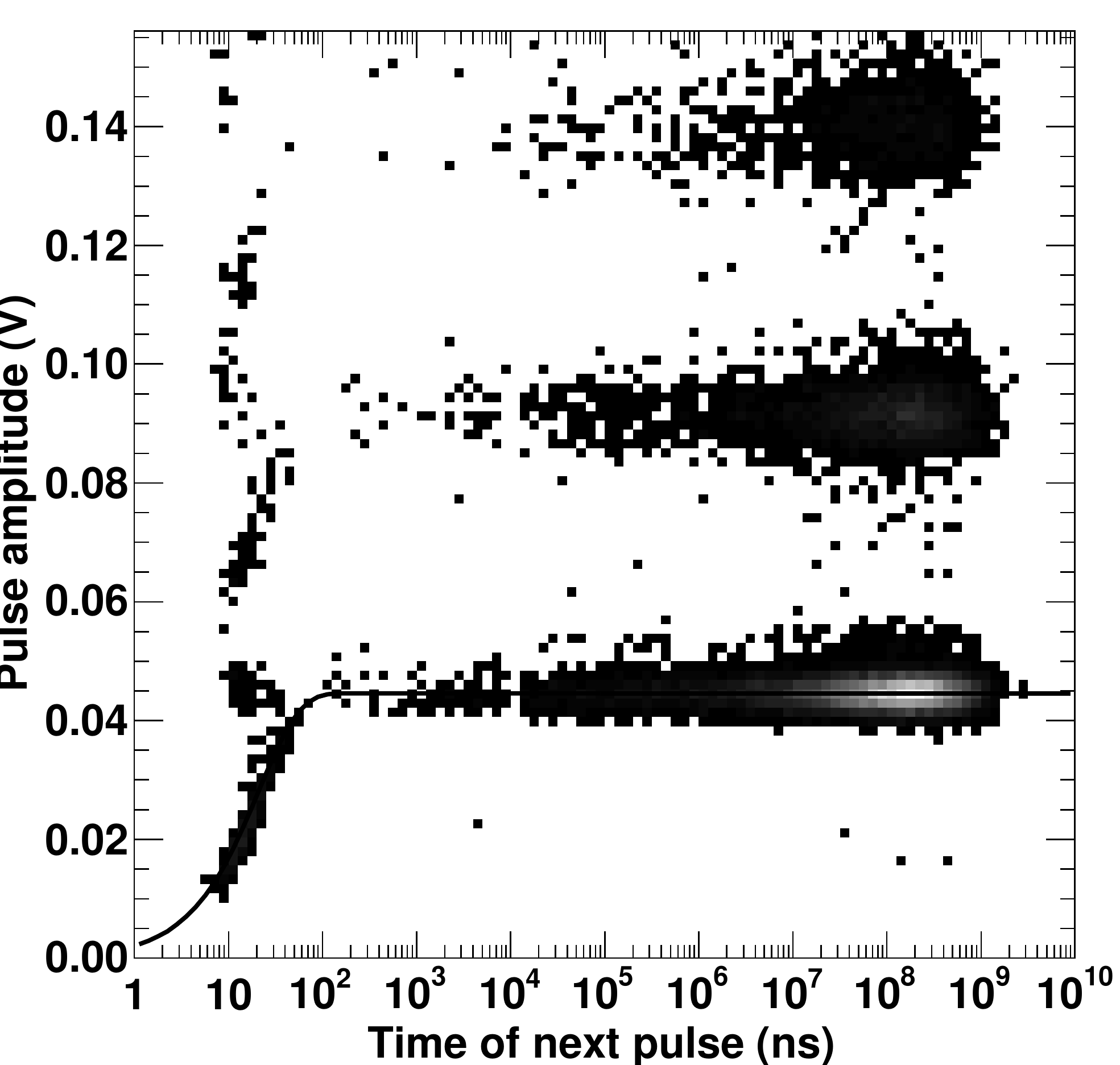}
 \caption{Statistical distribution of amplitude of the pulses following the trigger pulse as a function of the timing difference between this pulse and the trigger pulse for the Hamamatsu-VIS MPPC. The black line is the recovery function.}
 \label{fig:HPKDTimeAmp}
 \end{figure}

As mentioned earlier, the distribution of the timing difference between consecutive pulses is used to measure the dark noise and after-pulsing rates. The starting pulses are required to correspond to the oscilloscope trigger in order to properly account for the oscilloscope dead time. The starting pulses are also required to correspond to single pixel avalanches (i.e. triggers with cross-talk are excluded) in order to measure the after-pulsing rate generated by a single parent avalanche. On the other hand, the second pulse can have any amplitude (above the noise). Fig.~\ref{fig:HPKDTimeAmp} shows the amplitude of the second pulse as a function of the time difference with the trigger pulse. The main band at $\sim$40 mV corresponds to single pixel avalanches. Cross-talk yields pulses with amplitudes two, three, or more times larger. There is no correlation between time and amplitude except below 100 ns where a drop below the 40 mV value is clearly visible. After an avalanche the voltage across the diode indeed recovers with a time constant given by the product of the pixel capacitance and quenching resistance. The pulses that do not have lower amplitude even though they occurred within the recovery time scale must come from different pixels. It is likely that these pulses are delayed cross-talk, the seed charge carrier being created by photons from the parent avalanche subsequently diffusing to a neighboring pixel. This feature is also clearly visible for the FBK-2013 SiPM, while it is absent for KETEK. The recovery time constants are summarized in Table~\ref{tab:nuisanceSum}. The recovery time is an order of magnitude faster for the Hamamatsu-VIS MPPC.

  \begin{figure}
 \centering
\includegraphics[scale=0.45]{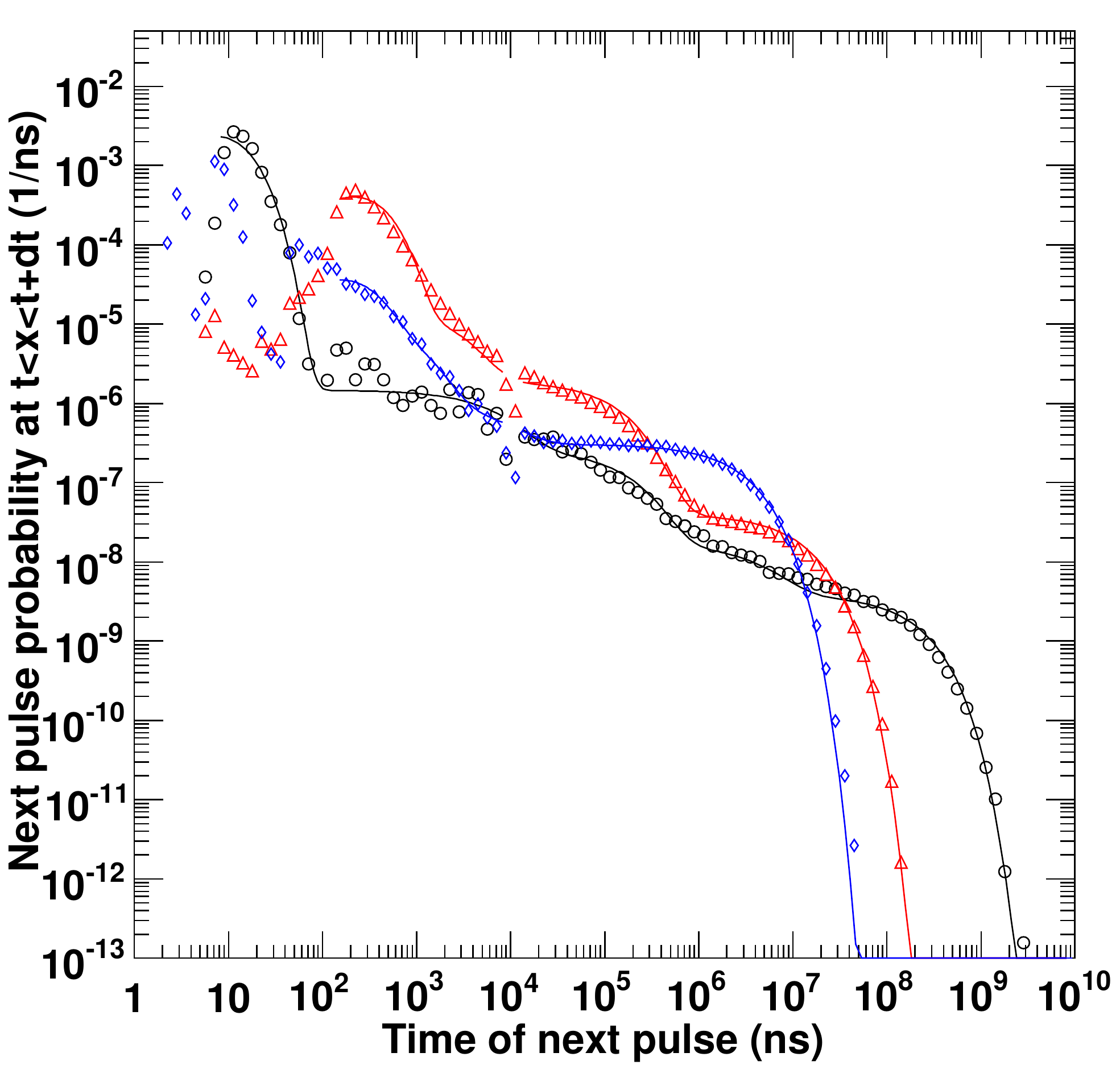}
 \caption{Probability that the next pulse occur between t and t+dt for the KETEK (red triangle), FBK-2013 (blue diamond) and Hamamatsu-VIS (black circle) SiPMs. The best fit functions are also shown. The gap in the distributions around 10$^4$ is due to the oscilloscope deadtime. Data error bars omitted for clarity.}
 \label{fig:DTime}
 \end{figure}

The timing distributions are shown in Fig.~\ref{fig:DTime} including the function that best fits the data. Each step on a log-log plot corresponds to a distinct exponential time dependence. The last step corresponds to distribution of dark pulses. It is clearly visible for all 3 devices, being largest for the FBK-2013 SiPMs and smallest for the Hamamamstu-VIS MPPC. The difference is striking because the FBK-2013 SiPMs are 9 times smaller in area than the KETEK and Hamamatsu-VIS devices. The dark noise rates per unit area are summarized in Table~\ref{tab:nuisanceSum}. Both KETEK and Hamamatsu-VIS devices are below the required 50Hz/mm$^2$ specification. But the FBK-2013 SiPM dark noise rate is an order of magnitude too high (a similar value was found for the FBK-2010 device). After-pulsing rates were also extracted by fitting the distributions in Fig.~\ref{fig:DTime}. Very slow after-pulses are visible and to some extent they could be accounted for in the dark noise rate. In table ~\ref{tab:nuisanceSum}, we included the total after-pulsing rate and the after-pulsing rate within the first 10 $\mu$s that is relevant for our application. The after-pulsing rate is low for both FBK-2013 and Hamamatsu-VIS SiPMs but it is high for KETEK devices, and exceeds the specification of 20\% for the combined rate of cross-talk and after-pulsing.

\subsection{Radio-assay of FBK SiPMs}
Table~\ref{tab:tefactiv} shows the concentrations of various elements in the FBK-2013 SiPMs as determined by NAA. 

\begin{table*}
     \caption{Concentrations of the 
     		elemental constituents determined by NAA for three FBK-2013 SiPM samples: June 2013, May 2014,
		and July 2014. The value given is the best fit result with Gaussian
		statical errors and 10\% systematic uncertainty associated with the calibration of the counting efficiency of the germanium detectors.
Where appropriate upper limits derived with the method outlined in~\cite{FC} are given.
    }
     \centering
   \begin{tabular}{c c   c  c }
       \hline
       \hline
       			&June 2013 irradiation							&May 2014 irradiation					&July 2014 irradiation\\\hline%
			& Concentration 								 & Concentration						& Concentration	 	\\
				\hline
       Na[ng/g]	&69.2\,$\pm$\,3.9\,$\pm$\,7.0						&1.56\,$\pm$\,0.09\,$\pm$\,0.16			&7.11\,$\pm$\,0.39\,$\pm$\,0.72 	\\ 
       K[ng/g]	&-1.5\,$\pm$\,10.8\,$\pm$\,0.1 ($<$17)				&-14.6\,$\pm$\,9.11\,$\pm$\,1.5	 ($<$4.8)		&3.81\,$\pm$\, 0.44\,$\pm$\,0.39		\\
       Sc[pg/g]	&47.8\,$\pm$\,2.8\,$\pm$\,4.8						&25.7\,$\pm$\,1.4\,$\pm$\,2.6				&7.33\,$\pm$\,0.40\,$\pm$\,0.74	\\   
       Cr[ng/g]	&1.25\,$\pm$\,0.08\,$\pm$\,0.13					&3.75\,$\pm$\,0.21\,$\pm$\,0.38			&0.396\,$\pm$\,0.022\,$\pm$\,0.040		\\
       Fe[ng/g]	&-											&34.4\,$\pm$\,1.9\,$\pm$\,3.5				&14.8\,$\pm$\,0.8\,$\pm$\,1.5		\\
       Co[pg/g]	&686\,$\pm$\,37\,$\pm$\,69						&137\,$\pm$\,73\,$\pm$\,14				&43.4\,$\pm$\,2.4\,$\pm$\,44		\\
       As[ng/g]	&288\,$\pm$\,14\,$\pm$\,29						&389\,$\pm$\,19\,$\pm$\,39				&590\,$\pm$\,29\,$\pm$\,60		\\
       Br[pg/g]	&8.91\,$\pm$\,0.46\,$\pm$\,0.90					&96.6\,$\pm$\,4.8\,$\pm$\,9.7				&169\,$\pm$\,9\,$\pm$\,17			\\
       Ti[$\mu$g/g]&(9.3\,$\pm$1.1\,$\pm$\,1.0$)\cdot$10$^{-3}$			&121\,$\pm$2\,$\pm$\,14					&38.3\,$\pm$\,1.6\,$\pm$\,3.9		\\
       Sb[pg/g]	&146$\pm$10	\,$\pm$\,15						&42.9$\pm$2.1	\,$\pm$\,4.3				&16.1\,$\pm$\,0.8\,$\pm$\,1.7	\\
       W[ng/g]	&-0.078\,$\pm$\,0.013\,$\pm$\,0.010 ($<$0.02)			&1.31\,$\pm$\,0.006\,$\pm$\ 0.14	 		&0.151\,$\pm$\,0.008\,$\pm$\,0.016	\\
       Au[pg/g]	&27.7\,$\pm$\,1.3\,$\pm$\,2.8						&69.0\,$\pm$\,3.3\,$\pm$\,6.9				&19.0\,$\pm$\,1.0\,$\pm$\,1.9	\\
       Ir[fg/g]	&10.4\,$\pm$\,33\,$\pm$\,1.1	($<$65)				&327\,$\pm$\,17\,$\pm$\,33				&16.1\,$\pm$\,1.8\,$\pm$\,1.7		\\
       Th[pg/g]	&2.2\,$\pm$\,2.3\,$\pm$\,0.3	($<$6.0)				&3.15\,$\pm$\,0.23\,$\pm$\,0.32			&0.17\,$\pm$\,0.12\,$\pm$\,0.02 (0.17$_{-0.11}^{+0.12}$)\\
       U[pg/g]	&-8\,$\pm$\,12\,$\pm$\,1 ($<$13)				&-7.18\,$\pm$\,4.21\,$\pm$\,0.72 ($<$9.1)		&1.0\,$\pm$\,6.1\,$\pm$\,0.1 ($<$10.9)	\\   %
       \hline \hline
     \end{tabular} 
     \label{tab:tefactiv}
        \end{table*}
It can be seen that the thorium concentration of all analyzed samples is no more than $\mathcal{O}$(10$^0$ pg/g). 
However, there is variation of a factor of ten in the amount of thorium seen in the samples. 
This may be related to the amount of backing that was on the chip. Further analysis is being planned to understand 
chip to chip variations. Our sensitivity to uranium is currently limited by the amount of arsenic contained in the chips, which is used as a dopant. The interference is due to the large number of $^{76}$As peaks and a similar half-life compared to $^{239}$Np. 

\section{Summary}
nEXO collaboration is investigating applicability of silicon photomultipliers for light detection in a large ultra-low background liquid xenon experiment. Here we report results of characterization of the SiPM samples obtained from several vendors. In particular, for the first time, the PDE at 175 nm is reported for FBK SiPMs. The device, based on a modified version of FBK's original n$^+$/p technology, shows an efficiency that approaches 10\% at $\sim$5V over-voltage, which is close to the value required by nEXO. We find that Hamamatsu devices optimized for light detection at 175 nm have PDE (\textit{in vacuo}) are close to the required 15\% at 2 V over-voltage, which is in a reasonable agreement with the measurements by the MEG collaboration~\cite{MEG:2014}. 
These results are encouraging for nEXO and other xenon experiments, which would benefit from using detectors capable of detecting scintillation light directly, without using wavelength shifters. 

Our measurements of the nuisance parameters show that cross-talk does not meet our specifications for the most recent available technologies of all three manufacturers. Hamamatsu MPPCs fulfill all other requirements. The tested FBK SiPMs show a too high dark noise rates, and KETEK SiPMs have also too high after-pulsing rates. Both Hamamatsu and FBK are working on improving the SiPM technology to address these issues.

Finally, a sensitive radio-purity assessment was performed with a large sample of SiPMs, indicating  levels of $^{40}$K, $^{232}$Th, and $^{238}$U contamination at or below $<$0.15, (6.9$\cdot$10$^{-4}$ - 1.3$\cdot$10$^{-2}$), and $<$0.11 mBq/kg, respectively. In spite of the observed variability between different samples, these results are very promising. 

\section*{Acknowledgment}
The authors would like to thank the University of Alabama machine shop and R.~Conley (SLAC National Accelerator Laboratory) for help with the machining and assembly of parts for the Stanford test setup; T.~Brunner and D.~Fudenberg for help with filling the scintillation sources with xenon gas; T.~Brunner, A.~Pocar, and L.~Yang for their helpful feedback on the manuscript, and the nEXO collaboration for discussions and advice. We thank Hamamatsu Photonics for providing samples of the MEG MPPCs and FBK for their continuous cooperation.

\bibliographystyle{IEEEtran}
\bibliography{SiPM_nEXO_IEEE}

\begin{thebibliography}{10}
\providecommand{\url}[1]{#1}
\csname url@samestyle\endcsname
\providecommand{\newblock}{\relax}
\providecommand{\bibinfo}[2]{#2}
\providecommand{\BIBentrySTDinterwordspacing}{\spaceskip=0pt\relax}
\providecommand{\BIBentryALTinterwordstretchfactor}{4}
\providecommand{\BIBentryALTinterwordspacing}{\spaceskip=\fontdimen2\font plus
\BIBentryALTinterwordstretchfactor\fontdimen3\font minus
  \fontdimen4\font\relax}
\providecommand{\BIBforeignlanguage}[2]{{%
\expandafter\ifx\csname l@#1\endcsname\relax
\typeout{** WARNING: IEEEtran.bst: No hyphenation pattern has been}%
\typeout{** loaded for the language `#1'. Using the pattern for}%
\typeout{** the default language instead.}%
\else
\language=\csname l@#1\endcsname
\fi
#2}}
\providecommand{\BIBdecl}{\relax}
\BIBdecl

\bibitem{Albert:2014b}
J.~B. Albert, D.~Auty, B.~E. Barbeau, P.S., D.~Beck, V.~Belov,
  C.~Benitez-Medina \emph{et~al.}, ``Search for {Majorana} neutrinos with the
  first two years of {EXO}-200 data,'' \emph{Nature}, vol. 510, pp. 229 -- 234,
  Jun 2014.

\bibitem{QVal:1998}
M.~Redshaw, E.~Wingfield, J.~McDaniel, and E.~G. Myers, ``Mass and
  double-beta-decay {$Q$} value of $^{136}\mathrm{Xe}$,'' \emph{Phys. Rev.
  Lett.}, vol.~98, p. 053003, Feb 2007.

\bibitem{n_silicon}
D.~T. Pierce and W.~E. Spicer, ``Electronic structure of amorphous {Si} from
  photoemission and optical studies,'' \emph{Phys. Rev. B}, vol.~5, pp.
  3017--3029, Apr 1972.

\bibitem{n_xenon}
V.~Solovov, V.~Chepel, M.~Lopes, A.~Hitachi, R.~F. Marques, and A.~Policarpo,
  ``Measurement of the refractive index and attenuation length of liquid xenon
  for its scintillation light,'' \emph{Nucl. Instrum. Meth. A}, vol. 516, no.
  2–3, pp. 462 -- 474, 2004.

\bibitem{Lorenzo:2014}
\BIBentryALTinterwordspacing
L.~Fabris, G.~De~Geronimo, S.~Li, V.~Radeka, and G.~Visser, ``Concepts of
  {SiPM} readout electronics,'' \emph{Workshop on large area, low background,
  VUV sensitive photo-detectors and associated electronics. IEEE Nuclear
  Science Symposium and Medical Imaging Conference, Seattle, WA}, 2014, (Date
  last accessed 6-July-2015). [Online]. Available:
  \url{http://info.ornl.gov/sites/publications/Files/Pub52995.pptx}
\BIBentrySTDinterwordspacing

\bibitem{fbk:2013}
N.~Serra, A.~Ferri, A.~Gola, T.~Pro, A.~Tarolli, N.~Zorzi, and C.~Piemonte,
  ``Characterization of new {FBK} {SiPM} technology for visible light
  detection,'' \emph{Journal of Instrumentation}, vol.~8, no.~03, p. P03019,
  2013.

\bibitem{MEG:2014}
W.~Ootani, K.~Ieki, T.~Iwamoto, D.~Kaneko, T.~Mori, S.~Nakaura \emph{et~al.},
  ``Development of {deep-UV} sensitive {MPPC} for liquid xenon scintillation
  detector,'' \emph{{Nucl. Instrum. Meth. A, \textnormal{in press, doi:
  http://dx.doi.org/10.1016/j.nima.2014.12.007}}}.

\bibitem{apd:2009}
R.~Neilson, F.~LePort, A.~Pocar, K.~Kumar, A.~Odian, C.~Prescott \emph{et~al.},
  ``Characterization of large area {APD}s for the {EXO}-200 detector,''
  \emph{Nucl. Instrum. Meth. A}, vol. 608, no.~1, pp. 68 -- 75, 2009.

\bibitem{omega}
http://www.omega.com.

\bibitem{ez}
http://www.ezag.com.

\bibitem{pelham}
http://www.pelhamresearchoptical.com.

\bibitem{cremat}
http://www.cremat.com.

\bibitem{root}
https://root.cern.ch/drupal.

\bibitem{Ham:v3}
{http://www.hamamatsu.com/resources/pdf/etd/PMT\_handbook\_v3aE.pdf}.

\bibitem{MAR6}
{http://www.minicircuits.com/pdfs/MAR-6SM+.pdf}.

\bibitem{Fit:2011}
A.~Vacheret, G.~Barker, M.~Dziewiecki, P.~Guzowski, M.~Haigh, B.~Hartfiel
  \emph{et~al.}, ``Characterization and simulation of the response of
  multi-pixel photon counters to low light levels,'' \emph{Nucl. Instrum. Meth.
  A}, vol. 656, no.~1, pp. 69 -- 83, 2011.

\bibitem{nudat}
{http://www.nndc.bnl.gov/nudat2/}.

\bibitem{Leonard:2008}
D.~Leonard, P.~Grinberg, P.~Weber, E.~Baussan, Z.~Djurcic, G.~Keefer
  \emph{et~al.}, ``Systematic study of trace radioactive impurities in
  candidate construction materials for {EXO-200},'' \emph{Nucl. Instrum. Meth.
  A}, vol. 591, no.~3, pp. 490 -- 509, 2008.

\bibitem{acid}
http://wwwsci.seastarchemicals.com.

\bibitem{NIST:2004}
https://www s.nist.gov/srmors/view\_detail.cfm?srm=1633b.

\bibitem{JENDL:2011}
K.~Shibata, O.~Iwamoto, T.~Nakagawa, N.~Iwamoto, A.~Ichihara, S.~Kunieda
  \emph{et~al.}, ``Jendl-4.0: A new library for nuclear science and
  engineering,'' \emph{Journal of Nuclear Science and Technology}, vol.~48,
  no.~1, pp. 1--30, 2011.

\bibitem{ootani:2014}
{http://www.hamamatsu.com/resources/pdf/etd/PMT\_TPMZ0001E01.pdf}.

\bibitem{FC}
G.~J. Feldman and R.~D. Cousins, ``Unified approach to the classical
  statistical analysis of small signals,'' \emph{Phys. Rev. D}, vol.~57, pp.
  3873--3889, Apr 1998.

\end{thebibliography}

\end{document}